%% file: SterileNeutrinoLG_Final.tex
\documentclass[usenatbib]{mnras}
\usepackage{graphicx, amssymb, aas_macros, natbib}
\usepackage{amsmath}
\usepackage{color}
\hypersetup{colorlinks=true, citecolor=blue, urlcolor=blue, linkcolor=blue}

\input{extra_preamble.tex}

\begin{document}

\title[Sterile neutrinos in the local and high-$z$ Universe]{Resonant Sterile Neutrino Dark Matter in the Local and High-$z$ Universe}

\author[B. Bozek et al.] 
{Brandon Bozek$^{1}$\thanks{bozek@astro.as.utexas.edu}, 
Michael~Boylan-Kolchin$^1$, 
Shunsaku~Horiuchi$^2$,\newauthor 
Shea~Garrison-Kimmel$^{3}$, 
Kevork~Abazajian$^4$, 
and James~S.~Bullock$^4$\\
$^1$Department of Astronomy, The University of Texas at Austin, 2515 Speedway, Stop C1400, Austin, TX 78712, USA\\
$^2$Center for Neutrino Physics, Department of Physics, Virginia Tech, Blacksburg, VA 24061, USA\\
$^3$TAPIR, California Institute of Technology, Pasadena, CA 91125, USA\\
$^4$Department of Physics and Astronomy, University of California at Irvine,
Irvine, CA 92697, USA}

\pagerange{\pageref{firstpage}--\pageref{lastpage}} 
\pubyear{2015}

\maketitle
\label{firstpage}

\begin{abstract} 
  Sterile neutrinos comprise an entire class of dark matter models that,
  depending on their production mechanism, can be hot, warm, or cold dark matter.  We
  simulate the Local Group and representative volumes of the Universe in a
  variety of sterile neutrino models, all of which are consistent with the
  possible existence of a radiative decay line at $\sim$3.5 keV. We compare
  models of production via resonances in the presence of a lepton asymmetry
  (suggested by Shi \& Fuller 1999) to ``thermal'' models.  We find that
  properties in the highly nonlinear regime -- e.g., counts of satellites and
  internal properties of halos and subhalos -- are insensitive to the precise
  fall-off in power with wavenumber, indicating that nonlinear evolution
  essentially washes away differences in the initial (linear) matter power
  spectrum. In the quasi-linear regime at higher redshifts, however,
  quantitative differences in the 3D matter power spectra remain, raising the
  possibility that such models can be tested with future observations of the
  Lyman-$\alpha$ forest. While many of the sterile neutrino models largely
  eliminate multiple small-scale issues within the Cold Dark Matter (CDM)
  paradigm, we show that these models may be ruled out in the near future 
  via discoveries of additional dwarf satellites in the Local Group.
\end{abstract}


\section{Introduction}
\label{sec:intro} 

The triumphs of the now-standard \lcdm\ model are manifold (see
\citealt{Frenk2012} and \citealt{Primack2015} for recent reviews). Nagging doubts about \lcdm\ on
small scales (corresponding to linear perturbations with masses of
$\la 10^{10}\,\msun$) have lingered, however, owing in large part to the 
difficulty of modeling low-mass perturbations from the linear to the nonlinear regime. 
In studying the highly nonlinear regime, a number of issues have arisen on these scales. In particular, several
possible conflicts exist between theory and observations on the scale of dwarf
galaxies ($\mstar \la 0.1\,L^{\star}$ or $\mhalo \la 10^{11}\,\msun$ at
$z=0$). These small-scale issues include: the over-prediction of the abundance
of low-mass dark matter halos relative to the observed number of satellite
galaxies in the Local Group (the Missing Satellites Problem --
\citealt{Klypin1999,Moore1999}), the prediction of a central cusp in the dark
matter halo density profile while the density profile of dwarf galaxies inferred
from observations favor a core (the Core-Cusp Problem -- \citealt{Flores1994,
  Moore1994}), and the over-prediction of the density in the subhalos expected
to host the bright, classical Milky Way dwarf galaxies (the Too Big To Fail
(TBTF) Problem -- \citealt{Boylan-Kolchin2011,Boylan-Kolchin2012}).

\begin{table*}
\begin{center}
\caption{WDM Models. {\it Column 1}: Model label; {\it Column 2}: Physical parameter that defines the model: cosmological lepton number ($L$), mixing angle between sterile and active neutrinos ($\sin^{2}(2\theta)$), or thermal WDM particle mass ($m$); {\it Column 3}: Half-mode mass; {\it Column~4}:~Was a Local Group pair simulated for this model?; {\it Column 5}: Final redshift of the cosmological volume simulation.}  
\label{tab:modellist}
\begin{tabular}{l | c | c | c | c }
\hline
Model & Physical Parameter & $M_{1/2} \, [\hmsun]$ & Local Group pair & $z_f$ \\ \hline \hline
L46 		& $L = 4.6\times10^{-4}$ 		& $2.1\times10^8$ & yes & 2 \\ 
THM2.9 	& $m = 2.9$~keV 	& $2.4\times10^8$ & no & 2 \\
S220 	& $\sin^{2}(2\theta) = 20\times10^{-11}$ & $2.2\times10^8$ & no & 2 \\  \hline
L7 		& $L = 7\times10^{-4}$ 		& $7.8\times10^8$ & yes & 0 \\ 
THM2 	& $m = 2$~keV 	& $9.4\times10^8$ & yes & 0 \\ 
S229 	& $\sin^{2}(2\theta) = 2.9\times10^{-11}$ & $8.2\times10^8$ & yes & 0 \\  \hline
L8 		& $L = 8\times10^{-4}$ 		& $2.1\times10^9$ & yes & 2 \\ 
THM1.6 	& $m = 1.6$~keV 	& $1.8\times10^9$ & no & 2 \\ 
S208 	& $\sin^{2}(2\theta) = 0.8\times10^{-11}$ & $1.9\times10^9$ & no & 2 \\  \hline
\end{tabular}
\end{center}

\end{table*}

Within the CDM paradigm, the resolution to these issues must lie in the baryonic
physics of galaxy formation. A combination of (global) photoionization and
stellar feedback may suppress star formation below some mass threshold, which
would resolve the Missing Satellites Problem \citep{Bullock2000,
  Kravtsov2004}. Proposed resolutions of the Core-Cusp problem and the TBTF
issue require supernova-driven gravitational potential fluctuations that remove substantial quantities of dark matter from the center of dark matter halos (possibly in concert with tidal stripping) to flatten cuspy density profiles into cores \citep{
  Governato2010, Brook2011, Governato2012, Zolotov2012,Di-Cintio2014,Onorbe2015}. 
  There is some debate in the literature as to whether these solutions can be generally viable \citep{ Penarrubia2012,
  Garrison-Kimmel2013}. In particular, the TBTF problem is also an issue for Local
Group and field halos, where stripping is unlikely to occur
\citep{Garrison-Kimmel2014a, Kirby2014, Papastergis2015}. The CDM small-scale
issues extending beyond the virial radii of the Milky Way (MW) and Andromeda
(M31) galaxies underscore the importance of studying the Local Group as a whole.

These small-scale issues in CDM have led to a reconsideration of Warm Dark
Matter (WDM) models of structure formation. In WDM scenarios, dark matter has a cosmologically-important
free-streaming scale that erases primordial perturbations below a characteristic
scale, one that is relevant for galaxy formation. For example, a $\sim 2$ keV 
Dark Matter (DM) particle produced in thermal equilibrium reduces power for (linear)
wavenumbers larger than $k \sim 10 \,\hmpcinv$, corresponding to (linear) mass
scales of $\sim 10^{9}\,\msun$ \citep{Bode2001,Schneider2012, Anderhalden2012, Anderhalden2013, Lovell2014}. Modifying the abundance of halos at this mass scale will
result in observable differences in the counts of dwarf galaxies around the
Local Group and in the field in the low-$z$ Universe.

keV-scale sterile neutrinos represent a plausible dark matter candidate from a
particle physics point-of-view \citep{Dodelson1994, Shi1999, Asaka2005, de-Gouvea2005}. The recent detection of a
possible line at 3.55 keV in the Perseus cluster \citep{Bulbul2014,Boyarsky2014},
stacked galaxy clusters \citep{Bulbul2014}, the Milky Way Galactic Center \citep{Boyarsky2015}, M31
\citep{Boyarsky2014}, and eight other clusters of galaxies at greater than 2$\sigma$ \citep{Iakubovskyi2015a}, while contentious, has further heightened interest in keV-scale dark matter: the radiative decay of a 7.1 keV sterile neutrino would
result in a 3.55 keV X-ray line. If this is indeed the origin of the line, it is the first non-gravitational signature of DM and is revolutionary.  However, even in this case, the effects of such a particle on structure formation
are \textit{not} uniquely determined by its mass: different production
mechanisms produce different momentum-space distributions of the sterile neutrino dark matter for particles of
identical mass.

WDM models have often been categorized in two types
(e.g. \citealt{Colombi1996}).``Early decoupled species'' such as gravitinos,
which are often referred to as ``thermal'' WDM, are produced in thermal
equilibrium but decouple early (at $T \gg 100\,{\rm GeV}$); for such particles,
the particle mass and decoupling temperature set the effect on large-scale
structure. ``Sterile neutrinos'' often is used to mean right-handed neutrinos
produced by oscillations produced at $T \lesssim 1\rm\ GeV$. So long as
$g_\ast$ is unchanged during their production (and in the absence of any
resonant production), one can derive a mapping between early decoupled species
(thermal WDM) and oscillation-produced sterile neutrino dark matter
\citep{Colombi1996, Viel2005}:
\begin{equation}
  m(\nu_{\rm s}) = 4.43\,{\rm keV}\,\left(\frac{m_{\rm thermal}}{1\,{\rm
       keV}}\right)^{4/3}\,\left(\frac{\Omega_{\rm DM}h^2} {0.1225}\right)^{-1/3}\, .
\label{eq:viel}
\end{equation}

In the well-known Dodelson-Widrow (\citeyear{Dodelson1994}) model of sterile
neutrinos produced via non-resonant oscillations, however, $g_\ast$
\textit{does} change during oscillation production. This change is particularly
large during the quark-hadron transition, which alters production significantly
\citep{Abazajian2002}. The applicability of Equation~\ref{eq:viel} to
cosmological interpretations of the effects of sterile neutrinos is therefore
not clear, a priori. \citet{Abazajian2006} calculated the effects of the change
of $g_\ast$ through the quark-hadron transition and the effects of the
commensurate neutrino opacity change. This results in a correction between
equivalent particle masses for thermal early-decoupled WDM and non-resonant
sterile neutrino WDM; the corrected relationship is given by
\begin{equation}
m(\nu_s) = 3.90\,{\rm keV} \left(\frac{m_{\rm thermal}}{1\rm\ keV}\right)^{1.294} \left(\frac{\Omega_{\rm DM}h^2}{0.1225}\right)^{-1/3}.
\label{eq:abaz}
\end{equation}
We have calculated this relation by matching the value of half-mode suppression between the non-resonant Dodelson-Widrow transfer functions calculated in \citet{Abazajian2006} and the thermal WDM transfer functions, e.g., in \citet{Viel2005}. Eq. \ref{eq:abaz} is accurate in the half-mode match to within $1\%$ given the two transfer function relations.
We note that while the difference in the normalization and
exponent in the thermal mass term of Eq. \ref{eq:abaz} relative to
Eq. \ref{eq:viel} is small, this difference can have a significant
(${\sim}20\%$) effect on Dodelson-Widrow sterile neutrino mass constraints,
particularly at larger particle masses that are currently being derived from
constraints based on Lyman-alpha forest data \citep{Baur2015}. The cosmological
effects of sterile neutrinos produced through early decoupling (e.g.,
\citealt{Patwardhan2015}) actually behave as thermal WDM in their
transfer function and not as oscillation-produced transfer
functions. 

In the case of the Shi-Fuller (\citeyear{Shi1999}) sterile neutrino model of
resonant production in the presence of a large lepton asymmetry, the momentum
distributions are substantially non-thermal and their transfer functions need to
be specifically calculated. The main effect of this non-thermal distribution is
to modify the linear theory transfer function relative to thermal models. How
this modification evolves in the nonlinear regime is not obvious and must be
studied with numerical simulations, in general. Furthermore, as shown by
\citet{Abazajian2014} and \citet{Venumadhav2015}, a large range of different
behaviors are possible for a Shi-Fuller sterile neutrino of a given mass, with
the mixing angle between sterile and ordinary neutrinos and cosmological lepton
number determining on which scales power is suppressed. Neither
Equation~\ref{eq:viel} nor Equation~\ref{eq:abaz} applies to Shi-Fuller sterile
neutrinos, therefore.

In this work, we explore a range of parameter space of the Shi-Fuller
(resonantly-produced) sterile neutrino DM model that can account for 3.55~keV
X-ray observations; our goal is to study the effects of such models on
cosmological structure formation. Previous studies of the MW in a WDM cosmology
focus only on thermal WDM particle models \citep{Lovell2012, Kennedy2014,
Lovell2014} or approximate resonant sterile neutrino DM with thermal WDM
transfer functions \citep{Bose2015}. Approximations are used since calculating the exact transfer
function for resonantly-produced sterile neutrino models, where particles are
produced out of thermal equilibrium, is challenging, and differences between
thermal and resonant sterile neutrino transfer functions appear small. Here, 
as was done in a companion work \citep{Horiuchi2015}, 
we use sterile neutrino transfer functions based on detailed Boltzmann code calculations and evaluate the relevance of
thermal WDM approximations used to place constraints on sterile neutrino model parameter space. 

The focus of this paper is two-fold. First, we examine the properties of Local
Group galaxies using zoom-in, dissipationless {\it N}-body simulations for a variety of
WDM models in order to explore solutions to CDM small-scale issues and predict
distinguishable features of a WDM Local Group. Second, we determine the
relevance of thermal WDM approximations to resonant sterile neutrino WDM models
by comparing Local Group properties, DM halo mass functions in large
cosmological volumes, and the 3D matter power spectrum at high-redshift.

The paper is organized as follows. In Section \ref{sec:Sims} we describe the WDM
models evaluated in this work, the numerical simulation details, and analysis
methods. We present our results for the Local Group
simulations in Section \ref{sec:LG} and field halo velocity functions of cosmological volume
simulations in Section \ref{sec:FieldHaloAbund}. We give the high-redshift matter power spectrum results of the cosmological volume simulations in Section \ref{sec:QuasiLinear}. In Section
\ref{sec:Conclusions} we discuss our results and present our conclusions.

\section{Simulations}
\label{sec:Sims} 

\begin{figure}
\begin{center}
\includegraphics[width=0.49\textwidth]{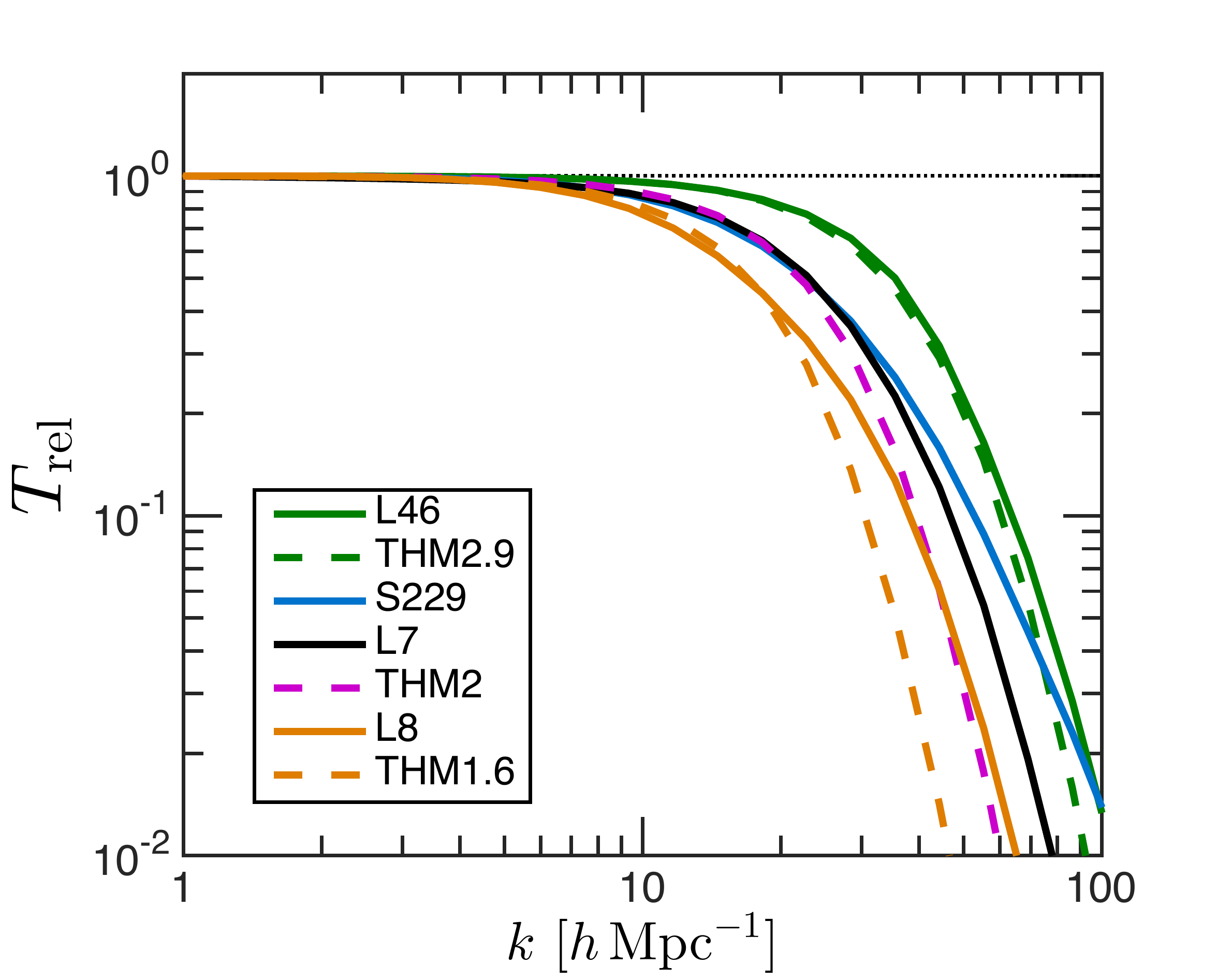}
\caption{Suppression of the linear matter power spectrum of resonantly-produced
  sterile neutrino models (solid lines) and their best-fit thermal equivalent
  model (dashed lines) relative to CDM, where
  $T_{\rm rel} = \sqrt{P_{\rm WDM}/P_{\rm CDM}}$. The L46 (solid; green), L7 (solid; black), and L8 (solid; orange) models are
  based on the sterile neutrino DM production calculations of \protect\cite{Abazajian2014}, while S229 (solid; blue) model is based on the
  more accurate treatment of \protect\cite{Venumadhav2015}. The L7 and S229 models both share an equivalent thermal WDM model of $m = 2$~keV (THM2/dashed; magenta). The shape and large-$k$ behavior of the WDM transfer functions vary among the sterile neutrino models and compared with their thermal equivalent models. 
\label{fig:trans_func}
}
\end{center}
\end{figure}

\begin{figure}
\begin{center}
\includegraphics[width=0.49\textwidth]{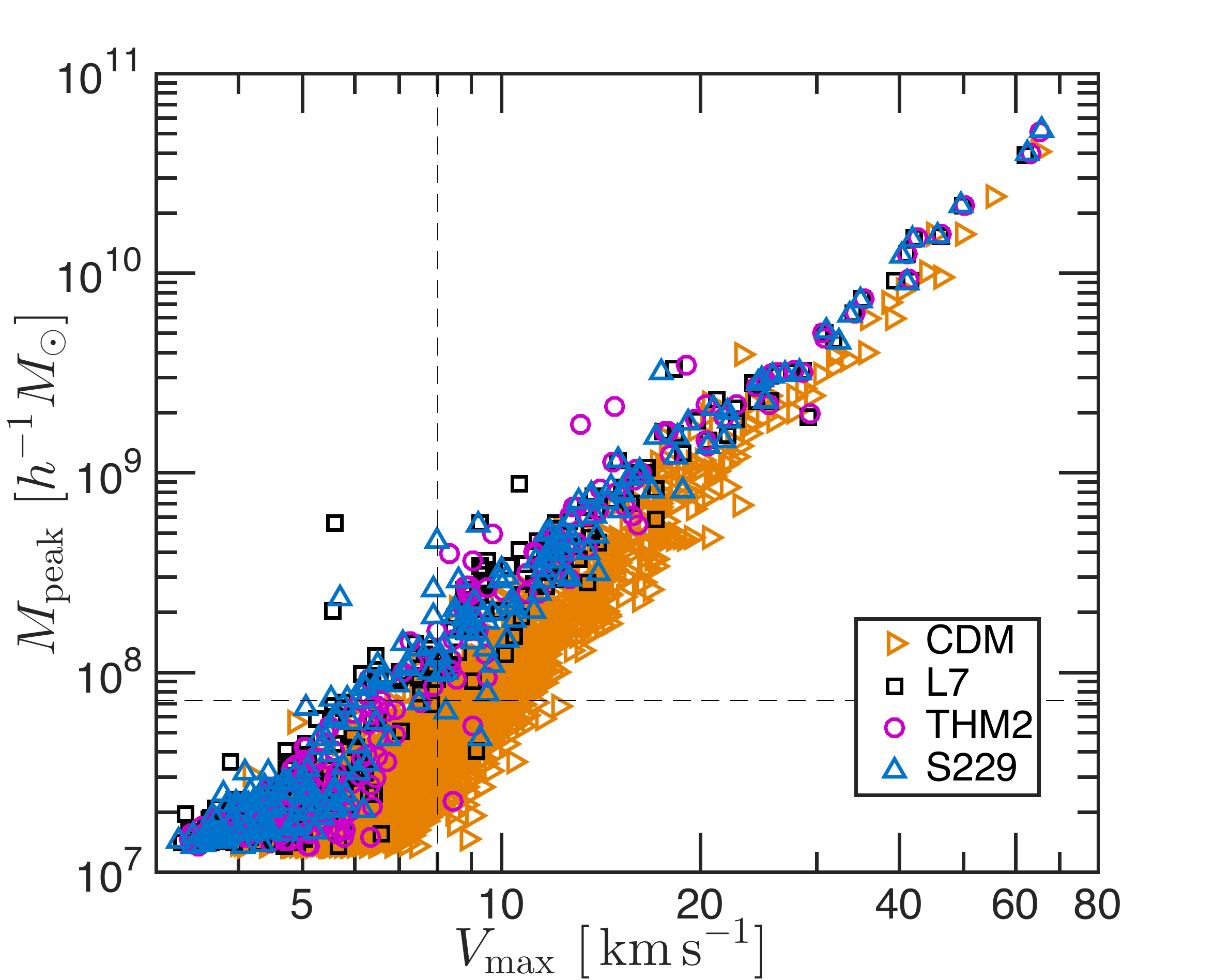}
\caption{The peak mass of CDM and WDM Local Group field halos with respect to $\vmax$ at $z = 0$. For a fixed
  peak mass bin, the WDM halos have a smaller $\vmax$ value, indicating a lower
  central density. The three WDM models shown here (L7, THM2, and S229) have the same half-mode suppression scale. Their Local Group field halo sets shown here have a similar distribution in
  $M_{\rm peak} - V_{\rm max}$ space.
  The vertical dashed-line shows the convergence limit of the
  simulation of $\vmax = 8 \, \kms$ and the horizontal dashed-line gives the
  limiting mass $M_{\rm lim}$ as defined by \protect\cite{Lovell2014}, below which
  the halos may have formed via artificial fragmentation. Only a few points fall
  in the bottom-right section as delineated by the dashed lines, indicating that
  our results are robust against discreteness effects.}
\label{fig:Mpeak_Relation}
\end{center}
\end{figure}

Throughout this work, we consider Shi-Fuller resonantly-produced sterile neutrino 
DM models with a particle mass of $m_{\nu} = 7.1$~keV that are consistent the 
observed $3.55$~keV line. As described in the Section \ref{sec:intro}, 
the particle mass does not uniquely specify the effect such a particle 
has on structure formation. We examine a range of sterile neutrino 
parameters that modify the linear matter power spectrum on scales relevant to small-scale structure formation. 
We additionally consider two calculations of Shi-Fuller models based on the work of Abazajian (\citeyear{Abazajian2014}; hereafter, A14) and also the more
recent results of Venumadhav (\citeyear{Venumadhav2015}; hereafter, V15).

The A14 models are parametrized by $L$, the lepton number of the Universe (relative to the photon number) prior to resonant production. We consider the models discussed in A14 of $L = 4.6\times10^{-4}, \,7\times10^{-4}, \,{\rm and}\,8\times10^{-4}$ (hereafter, L46, L7, and L8). These models bracket the region of parameter space that is consistent with producing the $3.55$~keV line and suppresses the linear matter power spectrum on small-scales that are relevant for DM halo counts, as it relates to the Missing Satellites Problem. Once $m_{\nu}$ is fixed and the lepton number is specified, the mixing angle $\sin^{2}(2\theta)$ is also determined. The case of L7 has a mixing angle of $\sin^{2}(2\theta) \approx 3\times 10^{-11}$, for example. 

\begin{table*}
\begin{center}
\caption{Host halo properties. {\it Column 1}: Model label; {\it Column 2}: Host halo virial mass; {\it Column 3}: Virial radius; {\it Column 4}: Maximum circular velocity; {\it Column 5} :Radius of circular velocity maximum; {\it Column 6}: Separation between host halos; {\it Column 7}: Number of halos above resolution limit; {\it Column 8}: Number of halos above the artificial fragmentation limit.}  \label{tab:hostprop}
\begin{tabular}{c | c | c | c | c | c | c | c}
\hline
Model 	& $M_{\rm vir}$	& $R_{\rm vir}$	&$V_{\rm max}$& $R_{\rm max}$ & $\Delta$ 	& $N(> 8\,\kms)$ & $N(>15\,\kms)$ \\ 
		& $[\hmsun]$ 	& $[\hmsun]$ 	& $[\kms]$ 		& $[\hkpc]$ & $[\hkpc]$ 		& & \\ \hline \hline
CDM (Thelma) 	& $1.05\times10^{12}$ & 256.69 & 171.48 & 61.69 & 623.7114 & 468 & 69 \\ 
CDM (Louise) 	& $8.50\times10^{11}$ & 239.52 & 158.82 & 33.07 & 623.7114 & 463 & 70 \\ \hline
L46 (Thelma) 	& $1.04\times10^{12}$ & 256.44 & 172.08 & 60.21 & 657.6297 & 101 & 38 \\ 
L46 (Louise) 	& $8.50\times10^{11}$ & 239.56 & 156.83 & 29.74 & 657.6297 & 106 & 37 \\  \hline
L7 (Thelma) 	& $1.04\times10^{12}$ & 256.30 & 171.03 & 60.81 & 665.3431 & 55 & 23 \\ 
L7 (Louise) 	& $8.47\times10^{11}$ & 239.22 & 153.69 & 25.26 & 665.3431 & 53 & 18 \\ \hline
THM2 (Thelma)& $1.05\times10^{12}$ & 256.76 & 170.99 & 61.23 & 663.9513 & 48 & 26 \\ 
THM2 (Louise) & $8.44\times10^{11}$ & 238.96 & 155.72 & 26.18 & 663.9513 & 48 & 17 \\  \hline
S229 (Thelma) & $1.05\times10^{12}$ & 256.62 & 170.89 & 65.4 & 661.1437 & 59 & 26 \\ 
S229 (Louise) 	& $8.49\times10^{11}$ & 239.39 & 154.65 & 25.36 & 661.1437 & 56 & 19 \\  \hline
L8 (Thelma) 	& $1.05\times10^{12}$ & 256.71 & 171.63 & 62.97 & 662.5239 & 36 & 21 \\ 
L8 (Louise) 	& $8.43\times10^{11}$ & 238.88 & 156.51 & 25.22 & 662.5239 & 30 & 11 \\  \hline
\end{tabular}
\end{center}
\end{table*}

The models from V15 include a more complete treatment of the
redistribution of lepton number and opacity of active neutrinos. We consider three V15 models 
parameterized by their mixing angles: $\sin^{2}(2\theta) = \{0.8, 2.9, \text{ and } 20\} \times 10^{-11}$ (hereafter, S208, S229, and S220). The V15 models are selected to bracket the region of sterile neutrino parameter space that is consistent with the candidate radiative decay line observations and have a similar half-mode suppression scale as the three A14 models. The lepton number (or equivalently here, the mixing angle) determines the resonantly-produced sterile neutrino momentum distribution and therefore the suppression of the linear matter power spectrum. For example, the L7 and S229 have similar values of $\sin^{2}(2\theta)$. These models are therefore comparable through their suppression of the linear matter power spectrum. Figure \ref{fig:trans_func} shows the `relative' transfer function, $T_{\rm rel} = \sqrt{P_{\rm WDM}/P_{\rm CDM}}$, that describes the suppression of the linear matter power spectrum relative to CDM for these models. The $T_{\rm rel}$ of the L7 and S229 models in Figure \ref{fig:trans_func} show a similar degree of suppression relative to CDM down to large values of $k$ where $T_{\rm rel} > 0.1$, albeit with a slightly different shape.

Each relative transfer function can be fit by a comparable relative transfer function for a thermal 
WDM model, which we subsequently refer to as its ``thermal equivalent''. The L7 and S229 cases have
a thermal equivalent model of $m = 2$~keV (THM2), while the L46 (L8) thermal equivalent
particle mass is higher (lower). The effective ``warmth'' of  the sterile neutrino models is 
then represented by the scale where the linear matter power spectrum 
diverges from CDM matter power spectrum, where warmer models suppress power on larger scales. 
Importantly, despite representing the best-fit case, the momentum distributions of 
the thermal equivalent models result in relative transfer functions that have different shapes from 
their sterile neutrino counterparts, as shown in Figure \ref{fig:trans_func}. We explore this point in detail throughout this work. 

The thermal equivalent transfer functions of
our fiducial L7 and S229 models match closely at the half-mode scale ($k_{1/2}$), where $T_{\rm rel}(k_{1/2}) = 0.5$,
as is shown in Fig.~\ref{fig:trans_func}. The half-mode scale indicates where free-streaming effects are significant, having suppressed the linear matter power spectrum by $50\%$ \citep{Avila-Reese2001}. The half-mode mass is defined as
\begin{equation}
  \label{eq:half_mode_mas}
  M_{1/2} = \frac{4 \pi}{3}\,\left(\frac{\lambda_{1/2}}{2}\right)^3\,\overline{\rho}\,,
\end{equation}
where $\lambda_{1/2} = 2 \pi / k_{1/2}$.

\begin{figure*}
\begin{center}
\begin{tabular}{c c }
\includegraphics[width=0.49\textwidth]{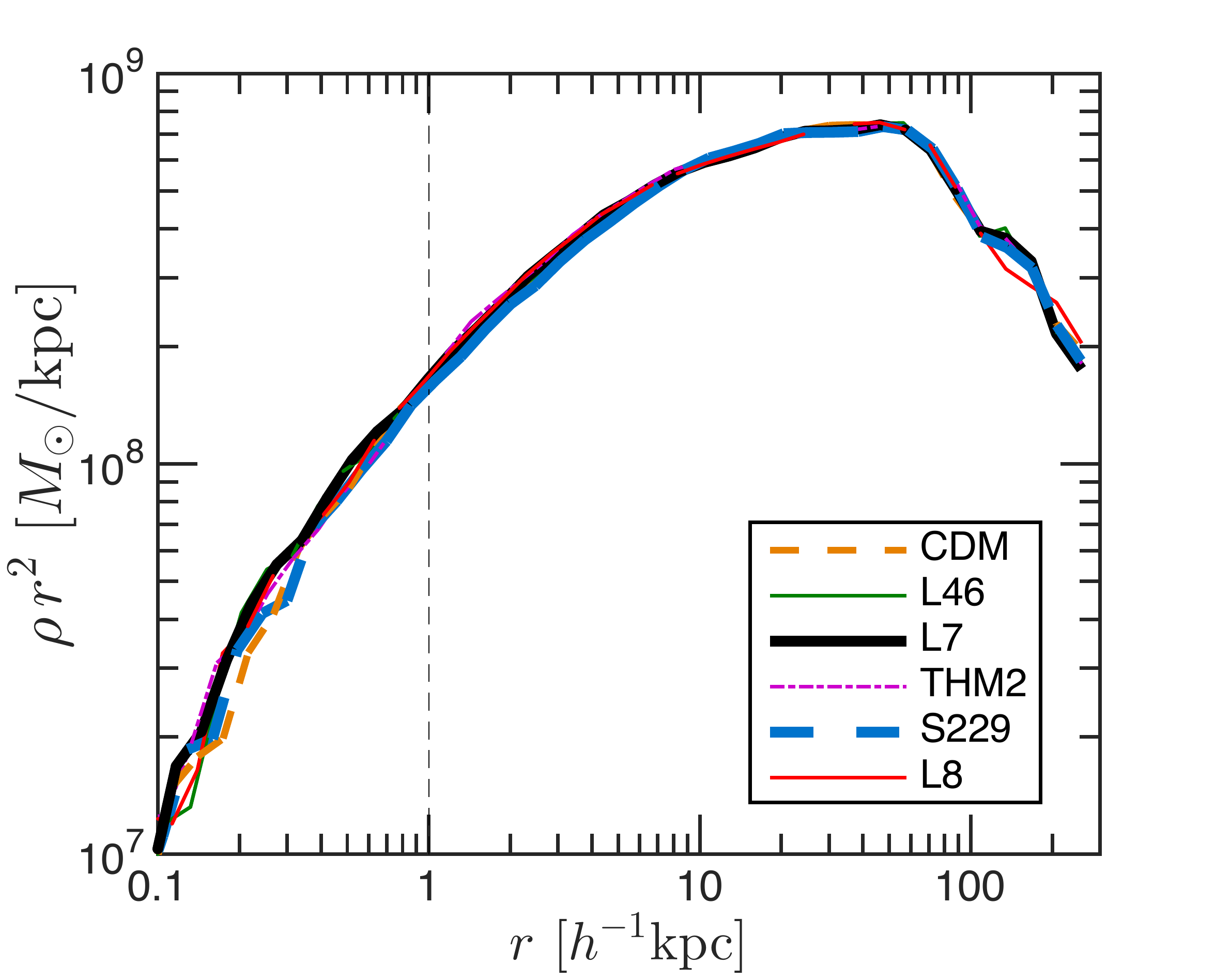} &
\includegraphics[width=0.49\textwidth]{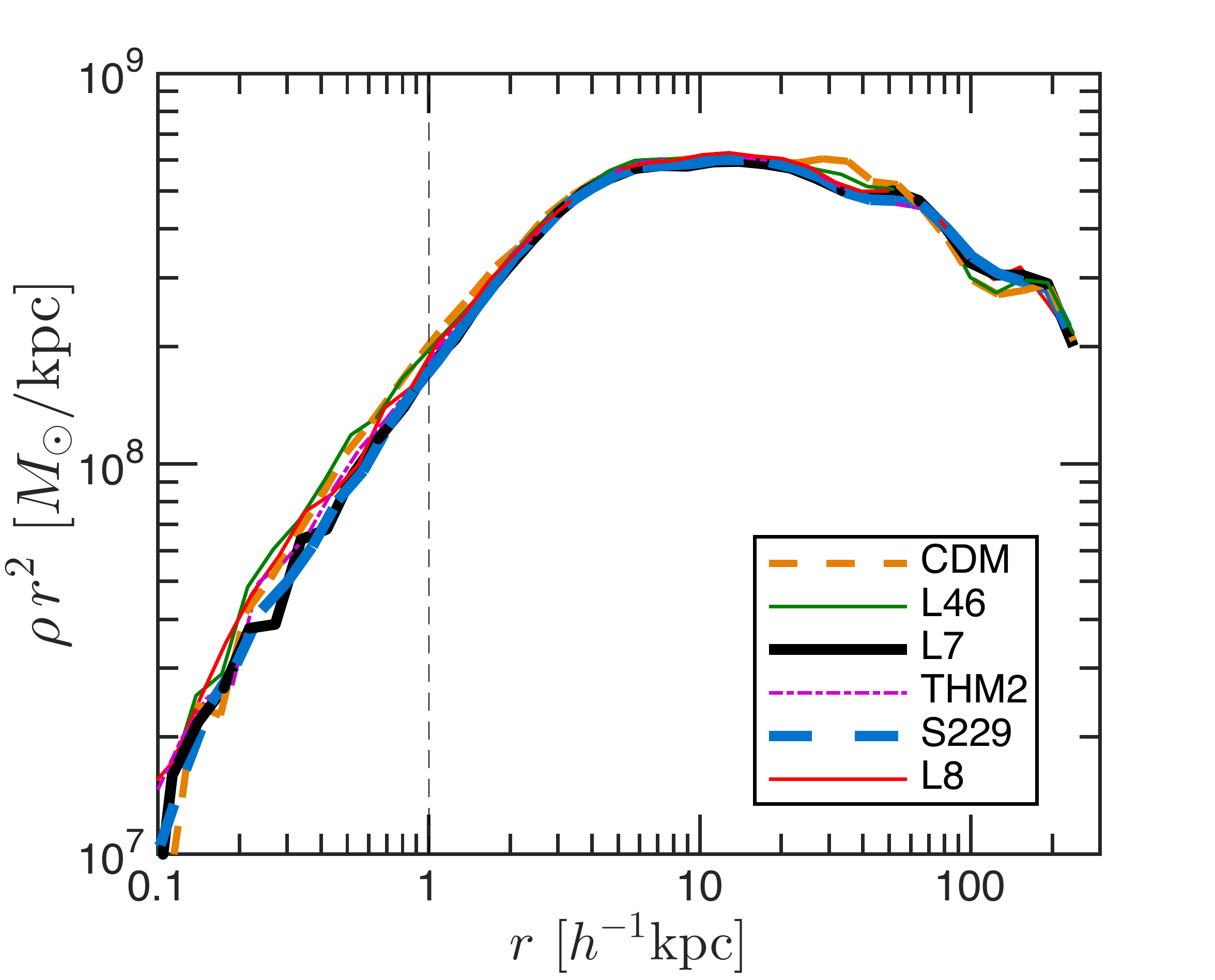} \\
\includegraphics[width=0.49\textwidth]{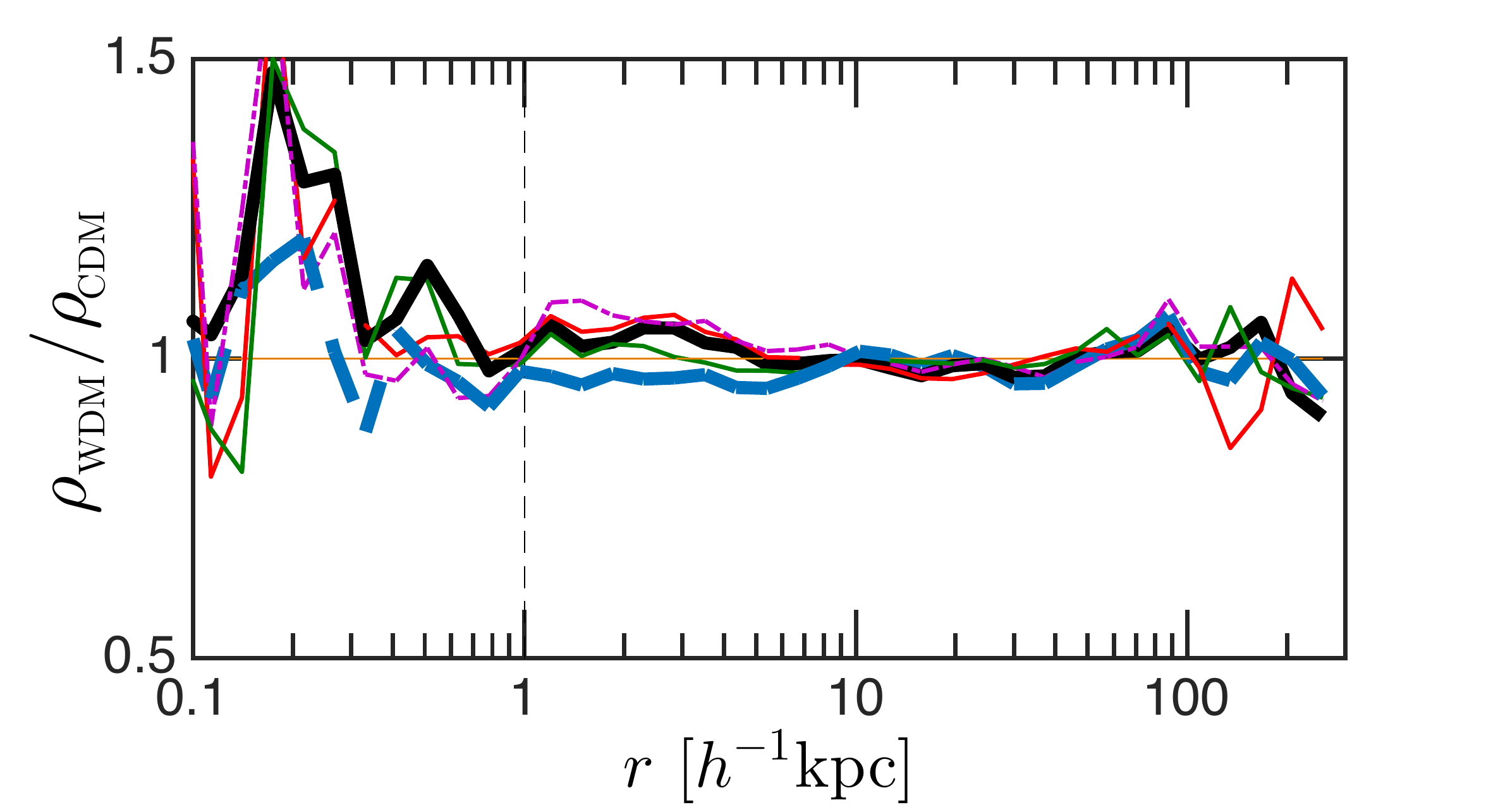} &
\includegraphics[width=0.49\textwidth]{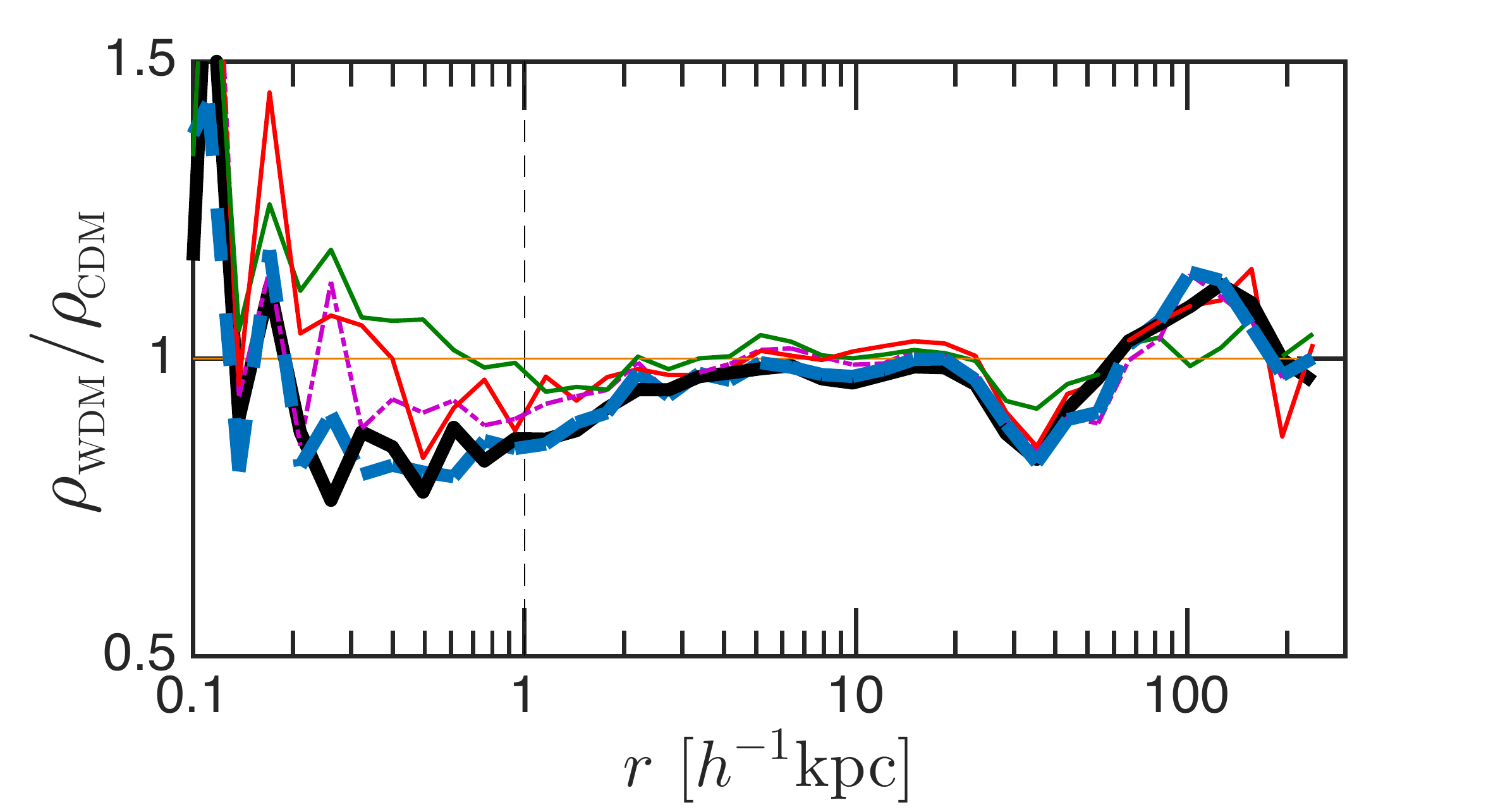} \\
\end{tabular}
\caption{The density profiles (top row) and ratio of WDM and CDM density
  profiles (bottom row) of host halos Thelma (left column) and Louise (right column)
  for the CDM (orange;dashed) and WDM models: L7 (black; thick-solid), THM2
  (magenta; dot-dashed), L46 (green; thin-solid), L8 (red; thin-sold), S229
  (blue; dashed). The dashed-vertical line marks the largest convergence radius
  ($r = 1\, \hkpc$) based on the criterion of \protect\cite{Power2003} as
  determined by the AHF halo finder \protect\citep{Knollmann2009}. Differences
  for $r < 1\, \hkpc$ are likely due to lack of numerical convergence. The density profiles for
  each model are very similar over the full radial range of each host. Small variations in the outer density profiles come from individual massive substructures.}
\label{fig:host_density}
\end{center}
\end{figure*}

We conduct two simulation suites. The first are a set of zoom-in, dark-matter-only {\it N}-body
simulations that are based on a single realization of a Local Group from the
Exploring the Local Volume In Simulations (ELVIS) suite
\citep{Garrison-Kimmel2014}. The ELVIS suite is a series of high-resolution
{\it N}-body simulations of 48 Milky Way-sized DM halos (24 in Local Group-like pairs,
24 isolated halos) in a \lcdm\ cosmology. The paired halos are selected to be
representative of the Local Group by resembling the Milky Way and M31 in mass,
separation, relative kinematics, and their ${\sim}2$~Mpc scale environment.  The ``Thelma and Louise'' Local Group
from the ELVIS suite was chosen to be the representative Local Volume targeted
in this work. The Local Group zoom-in simulations are run for four of the sterile
neutrino WDM models discussed above (L46, L7, L8, and S229) and a thermal
$m = 2$~keV WDM model. These simulations are compared to the original CDM simulation 
from the ELVIS suite. The three A14 models are selected to bracket a range of 
linear matter power suppression, as shown in Figure~\ref{fig:trans_func}, that is predicted to 
resolve the Missing Satellites Problem. The S229 and THM2 models, as discussed above, are
directly comparable to the L7 model and facilitate an exploration of the relative transfer function shape on Local Group properties.
\footnote{A related work by \citet{Horiuchi2015} compares zoom-in simulations of an isolated MW-sized DM halo for the V15 models S208, S229, and S220 with observations of the Local Group satellites specifically to determine sterile neutrino model resolutions to the Missing Satellites and Too Big To Fail problems.}
 
The Tree-PM $N$-body code Gadget-3, an updated version of the public Gadget-2 code
\citep{Springel2005a}, was used to perform the simulations, which assume a WMAP7
cosmology:
$\Omega_m = 0.266, \sigma_{8} = 0.801, \Omega_{\Lambda} = 0.734, n_s = 0.963,
\text{and } h = 0.71$
\citep{Larson2011}. The particle mass of the high-resolution zoom-in region is
$m_p = 1.3\times10^5 \hmsun$, which gives an effective resolution of $4096^3$
for the high-resolution region. The simulations start at an initial redshift of
$z_i = 125$ and employ a Plummer-equivalent softening of $\epsilon = 1\, \hkpc$,
which is constant in comoving units until $z = 9$, where it becomes fixed at $\epsilon = 100\, \hpc$
(physical) to redshift zero. The initial conditions for the simulation
were set-up using the MUSIC code \citep{Hahn2011}. 

\begin{figure*}
\begin{center}
\includegraphics[width=0.49\textwidth]{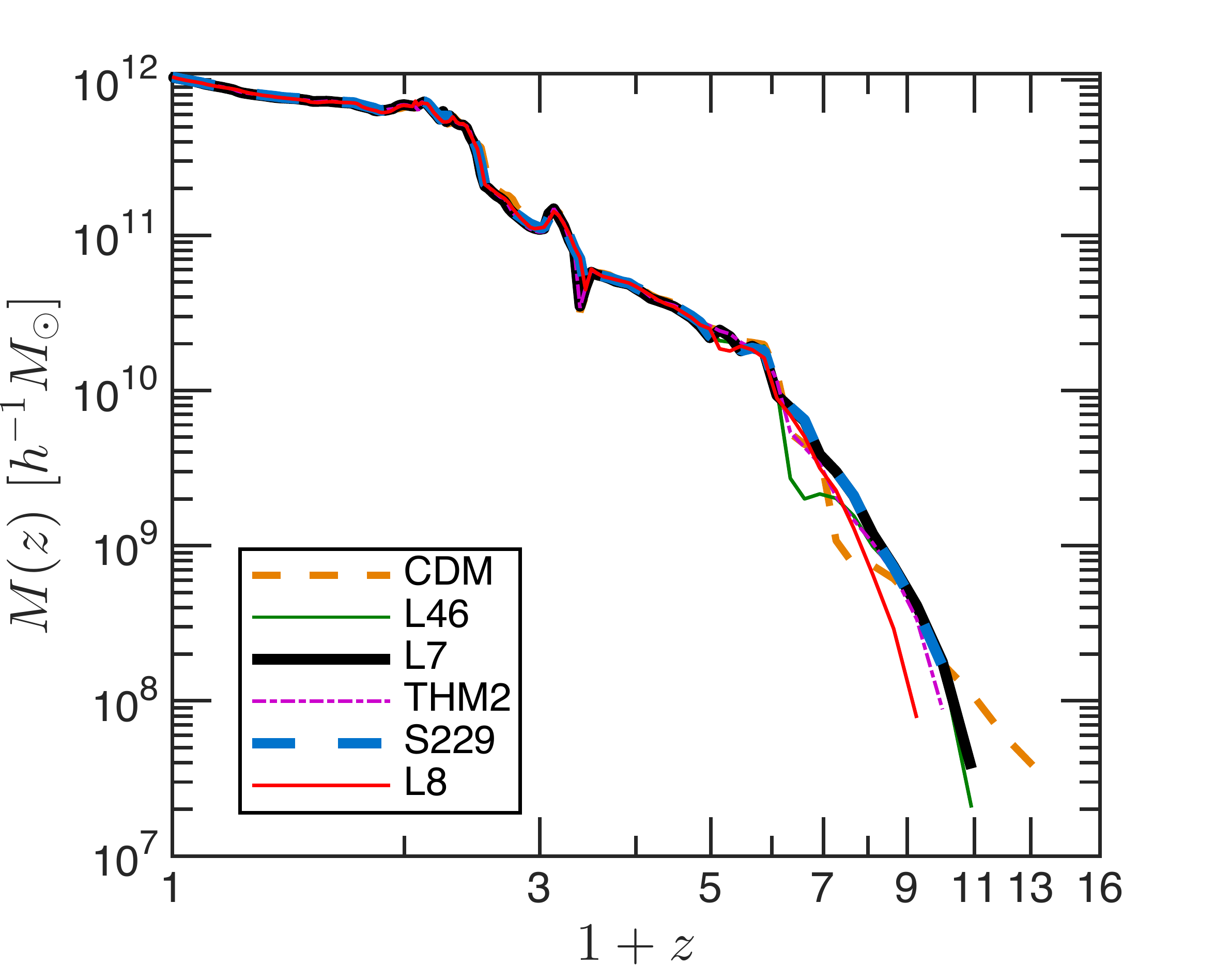}
\includegraphics[width=0.49\textwidth]{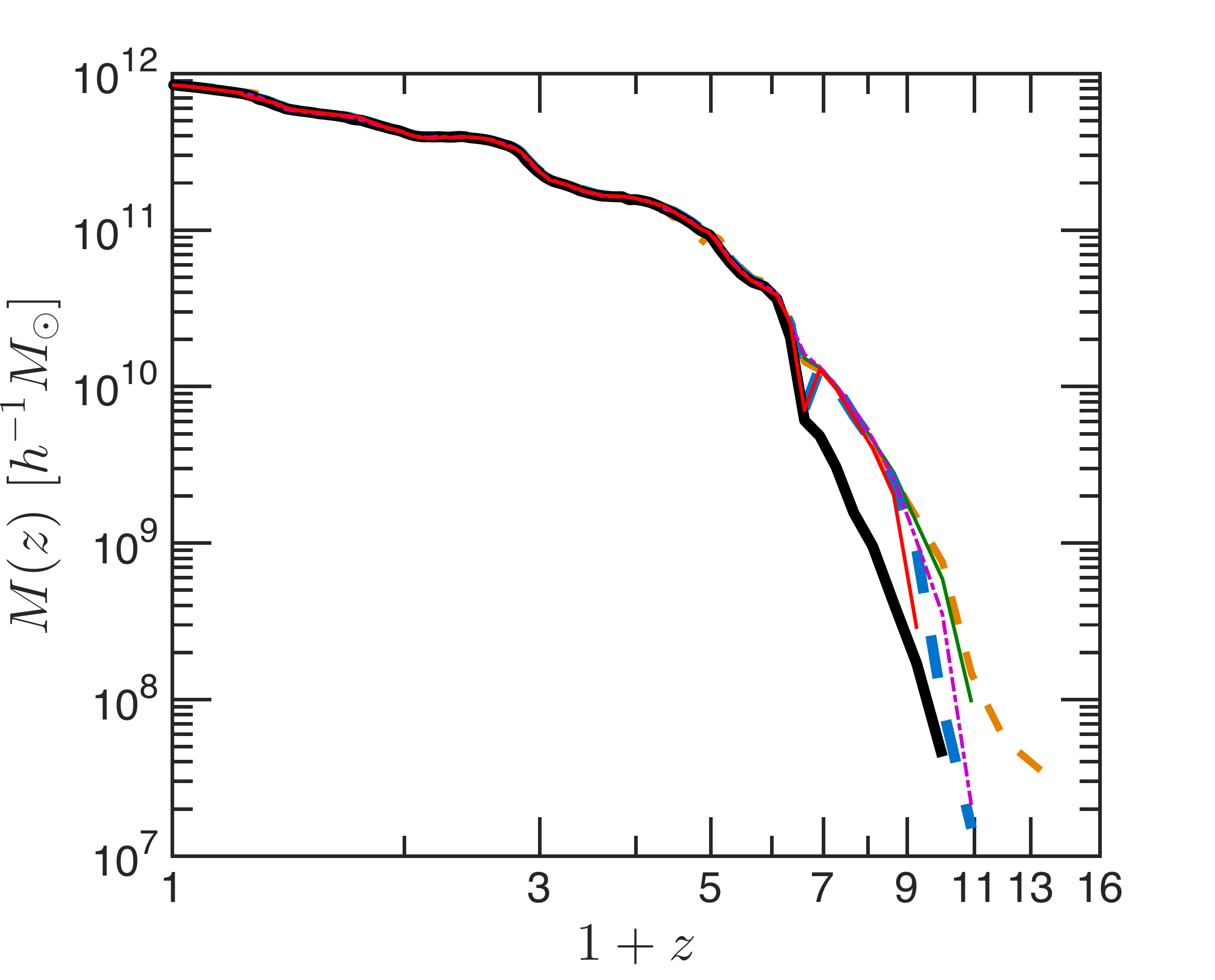}
\caption{The mass assembly histories of host halos Thelma (left panel) and Louise
  (right panel) for the DM models: L7
  (black; thick-solid), THM2 (magenta; dot-dashed), L4.6 (green; thin-solid), L8
  (red; thin-sold), S229 (blue; dashed), CDM (orange; dashed). After a redshift of $z = 5$ the virial mass evolution of each halo is identical for each DM model. Prior to $z=5$, there are small departures from a common assembly history that are not correlated with WDM model warmth. The host halos masses at initial collapse are near (or below) the half-mode mass of each WDM model (listed in Table \ref{tab:modellist}). The random departures from the early common assembly history are likely related to the mass build-up of halos near the different models' half-mode scales, where free-streaming effects are relevant.}
\label{fig:host_assem}
\end{center}
\end{figure*}

Our second suite of simulations are homogeneously-resolved, cosmological volume
dark-matter-only simulations of $N_p = 512^3$ particles in cubic volumes of box
length $L = 25 \, \hmpc$ with a WMAP7 cosmology. We simulate the A14 (L46,
L7, L8) and V15 (S220, S229, S208) sterile neutrino WDM models and their
best-fit thermal WDM models ($m = 1.6, 2, \text{and } 2.9$~keV; hereafter, THM1.6, THM2, and THM2.9). The Plummer-equivalent force softening for these simulations, $\epsilon = 1.25\,\hkpc$, is constant in comoving units for the duration of the simulation. The simulations start at $z_i = 125$. The L7, THM2, S229, and CDM simulations were evolved to $z=0$; the L46, L8, S208, and S220 models were simulated
to understand halo properties at high redshift and were therefore terminated at $z=2$. The nonlinear matter power spectrum of each simulation was calculated using the ProCorr code \citep{Obreschkow2013}.

We use the Amiga Halo Finder (AHF) \citep{Knollmann2009} to identify self-bound
dark matter halos. The virial mass of a dark matter halo, $M_{\rm vir}$, is defined as
the mass contained within a sphere out to the virial radius, $R_{\rm vir}$, where the
density of the halos is 200 times the mean density of the universe. We define
$R_{\rm max}$ as the radius where the maximum velocity, $\vmax$, of the circular
velocity profile, $V_{\rm c}(r) = \sqrt{GM(<r)/r}$, is reached for a dark matter halo
at redshift zero. The assembly histories of the dark matter halos are determined
by the MergerTree code that is included in AHF. The maximum circular velocity of
a dark matter halo over its entire assembly history is defined as $V_{\rm peak}$. Halo
partners across different simulations are also identified using the AHF
MergerTree code by finding the corresponding halo that shares the most particles
with the same Gadget-3 Particle IDs. For example, Halo A in the CDM simulation
is considered a partner halo of Halo B in the L7 WDM simulation if both halos
maximize the function $N_{AB}^2/(N_AN_B)$, where $N_{AB}$ is the number of
shared particles in each halo, $N_A$ is the number of particles in halo A, and
$N_B$ is the number of particles in halo B. This criterion is determined for
halos in each simulation type with respect to every other simulation type
and then compared for overlap between simulation types. When comparing halos 
directly across different simulation types, we do not include halos that fail this criterion, even
if the halos satisfy a separate selection criteria (such as a range of halo masses).  

\begin{figure*}
\begin{center}
\begin{tabular}{c c }
\includegraphics[width=0.49\textwidth]{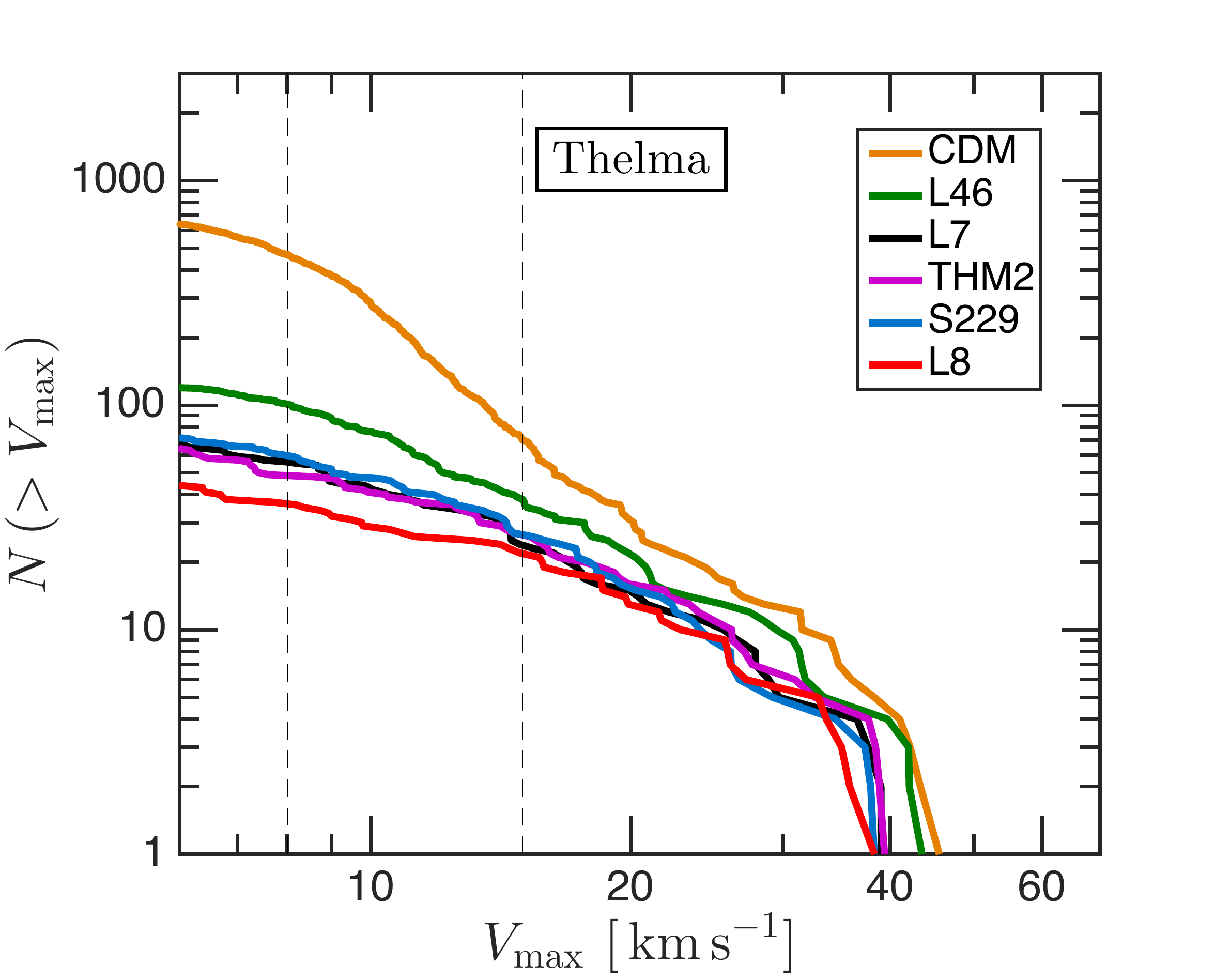} &
\includegraphics[width=0.49\textwidth]{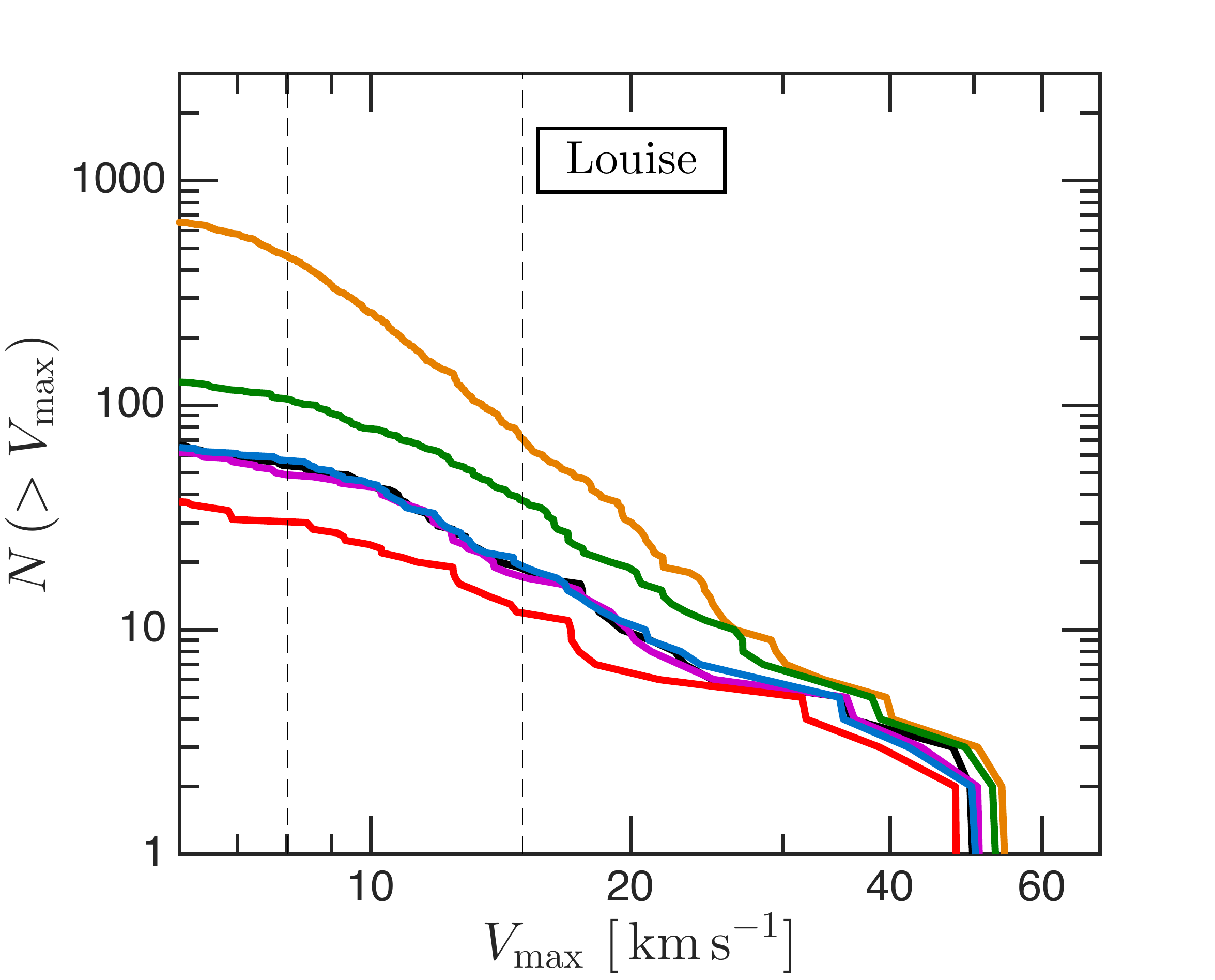} \\
\includegraphics[width=0.49\textwidth]{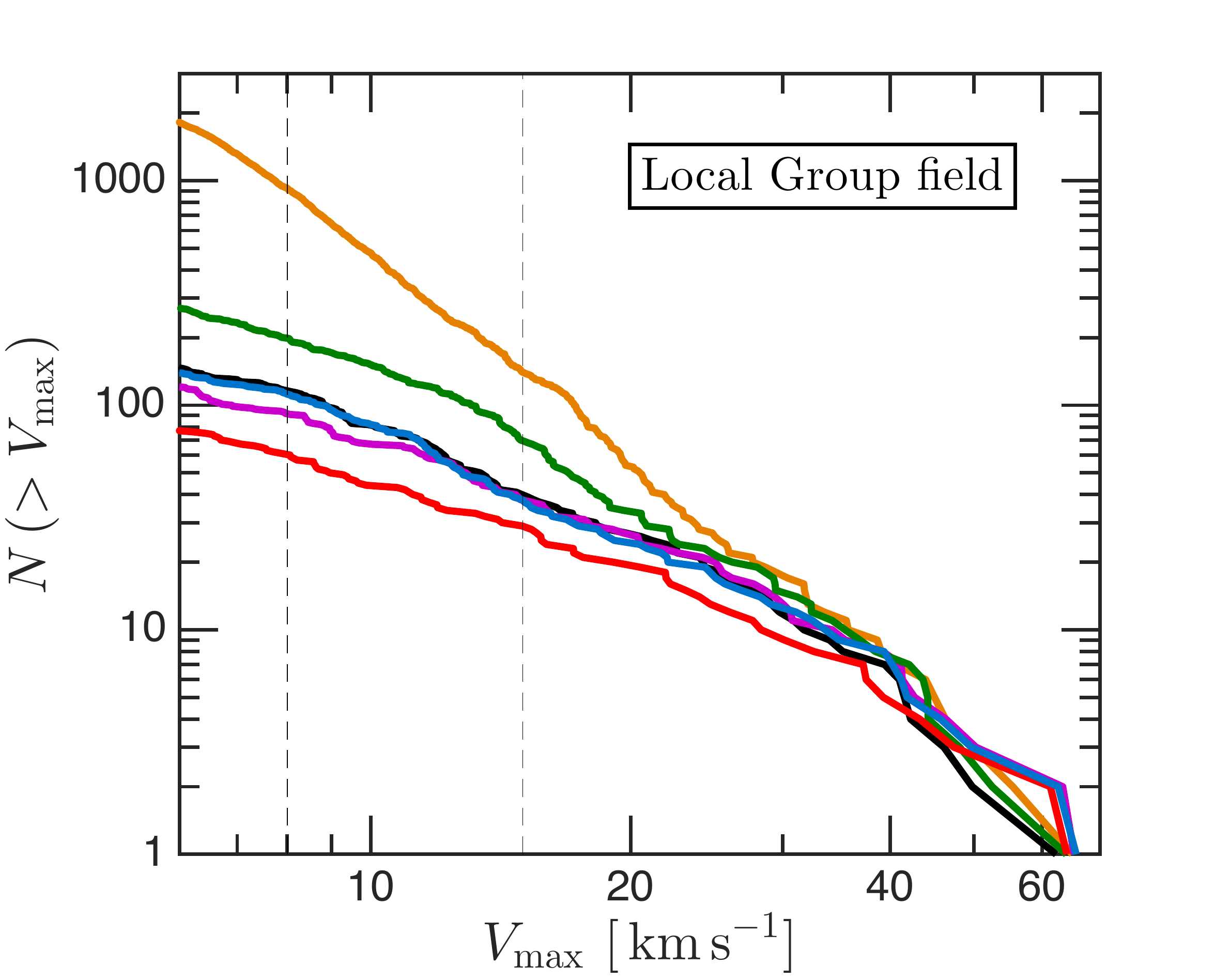} &
\includegraphics[width=0.49\textwidth]{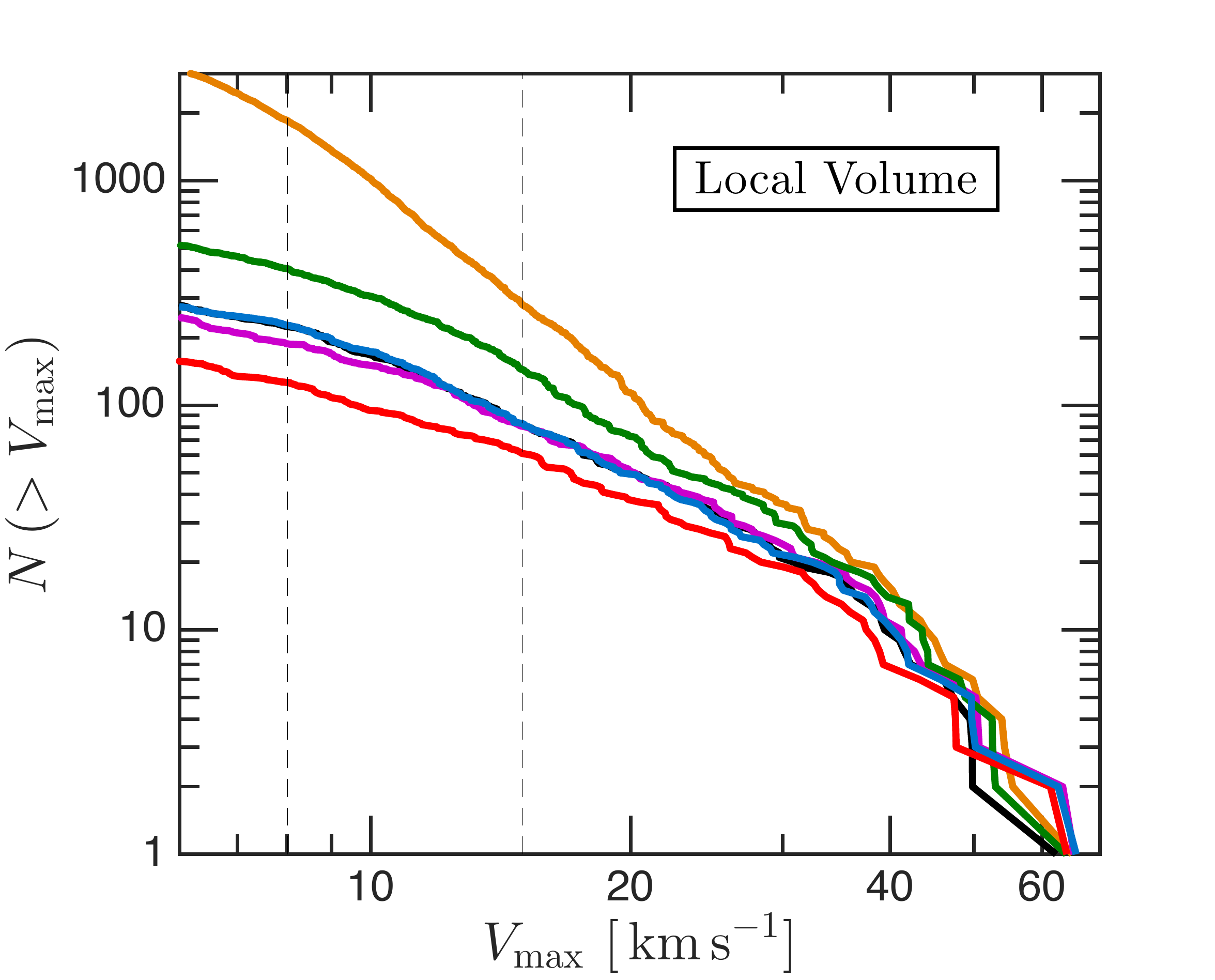} \\
\end{tabular}
  \caption{The cumulative subhalo $\vmax$ function for hosts Thelma (top left
    panel) and Louise (top right panel), the Local Group field halos (bottom
    left panel), and all Local Volume halos including host subhalos and field
    halos (bottom right panel) for each DM model. The convergence limit of
    $\vmax = 8 \, \kms$ and artificial fragmentation limit of $\vmax = 15\, \kms$ are respectively represented 
    by the dashed-vertical lines. The five WDM
    models suppress the number of subhalos at each mass relative to CDM.}
\label{fig:host_vmaxfunc}
\end{center}
\end{figure*}

The formation of halos below the free-streaming scale in WDM should be highly
suppressed; however, WDM numerical simulations find an excess of low-mass halos
compared to expectations both from theory and in comparison to CDM, where the CDM 
linear matter power spectrum has significantly more power than WDM. These halos in WDM are
numerical artifacts resulting from fragmentation of filaments induced by
discreteness noise and exist whether the initial particle load is grid-based or
glass-based \citep{Wang2007}. Two methods that address this issue of
spurious halos are: removing halos below a limiting mass and shape criterion as defined by
\cite{Lovell2014}, or considering only halos with a mass far above the mass
scale where spurious halo formation is a significant problem. In this work, we 
take the latter approach.

We note, however, that the contamination from spurious halos is likely to be
small even at the resolution limit of our Local Group simulation
($\vmax > 8 \,\kms$). Figure \ref{fig:Mpeak_Relation} shows the relationship of
the peak mass, $M_{\rm peak}$ to $\vmax (z=0)$ of Local Group field halos in the CDM, L7, S229, and THM2 
simulations. The vertical and horizontal dashed lines show the resolution limit
at $\vmax = 8 \, \kms$ and the limiting mass, defined as
$M_{\rm lim} = 10.1\, \bar{\rho} \, d\,k_{\rm peak}^{-2}$ \citep{Wang2007}, respectively. We 
follow \cite{Lovell2014} and multiply this value by $\kappa = 0.5$ in order to
best account for the mass scale below which artificial fragmentation dominates
to obtain a value of $M_{\rm lim} = 7.25\times10^7 M_{\odot}/h$. Halos in the bottom
right quadrant of Figure \ref{fig:Mpeak_Relation} are halos that are above our resolution limit, but below the
limiting mass, where their formation may have been the result of artificial
fragmentation. The small number of points in this section of the figure illustrates that our
results are not strongly contaminated by artificial fragmentation. However, when considering 
the internal properties of DM halos, we base our conclusions on halos with $\vmax > 15 \, \kms$, which we define as our ``artificial fragmentation limit''. This is a very conservative choice that selects halos with an  $M_{\rm peak}$ value significantly above $M_{\rm lim}$, as shown in Figure \ref{fig:Mpeak_Relation}. We quote results for halos with $\vmax > 8 \, \kms$ where appropriate.

\section{The Local Group}
\label{sec:LG}
\subsection{Local Group host halos}

\begin{figure*}
\begin{center}
\includegraphics[width=0.49\textwidth]{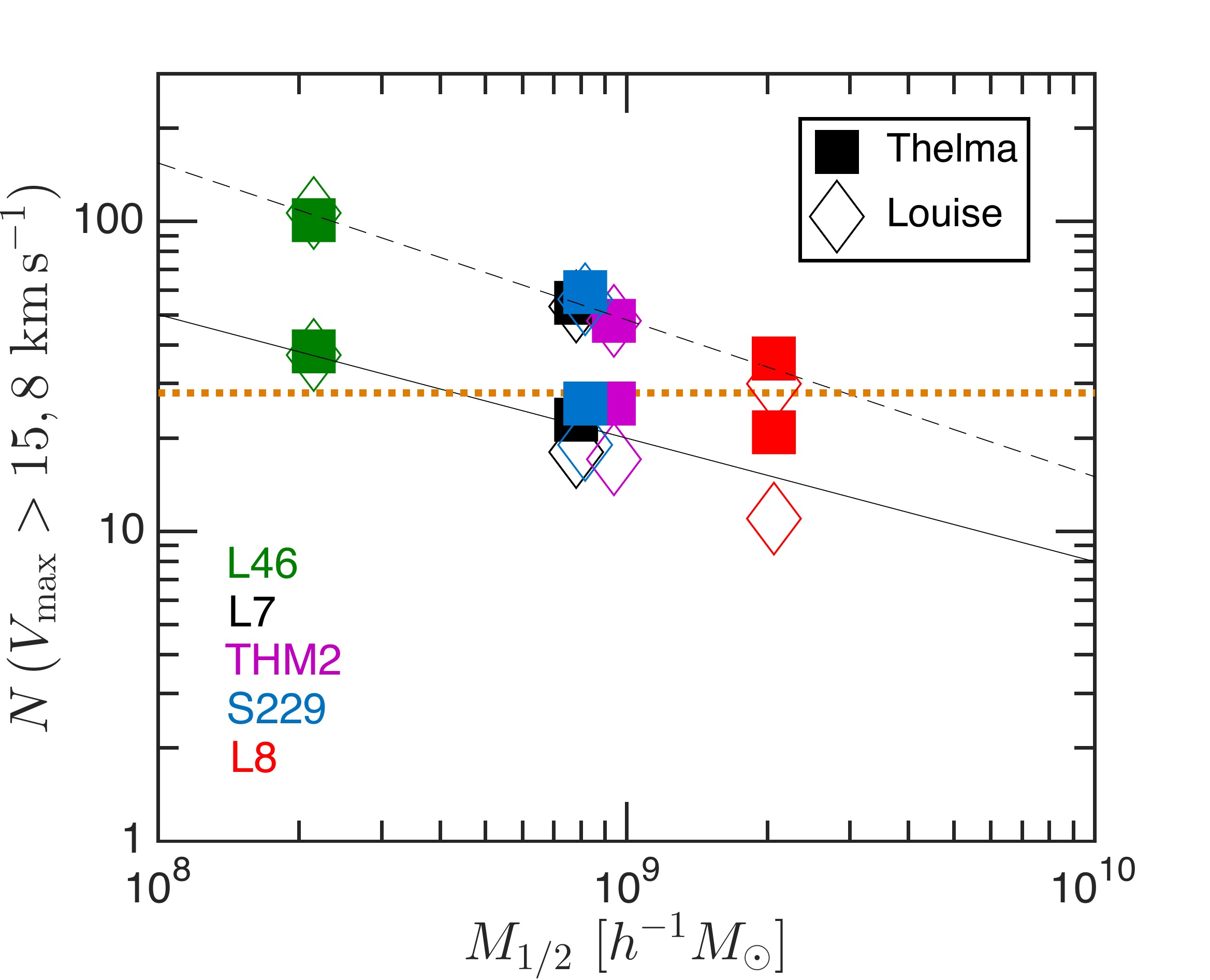}
 \includegraphics[width=0.49\textwidth]{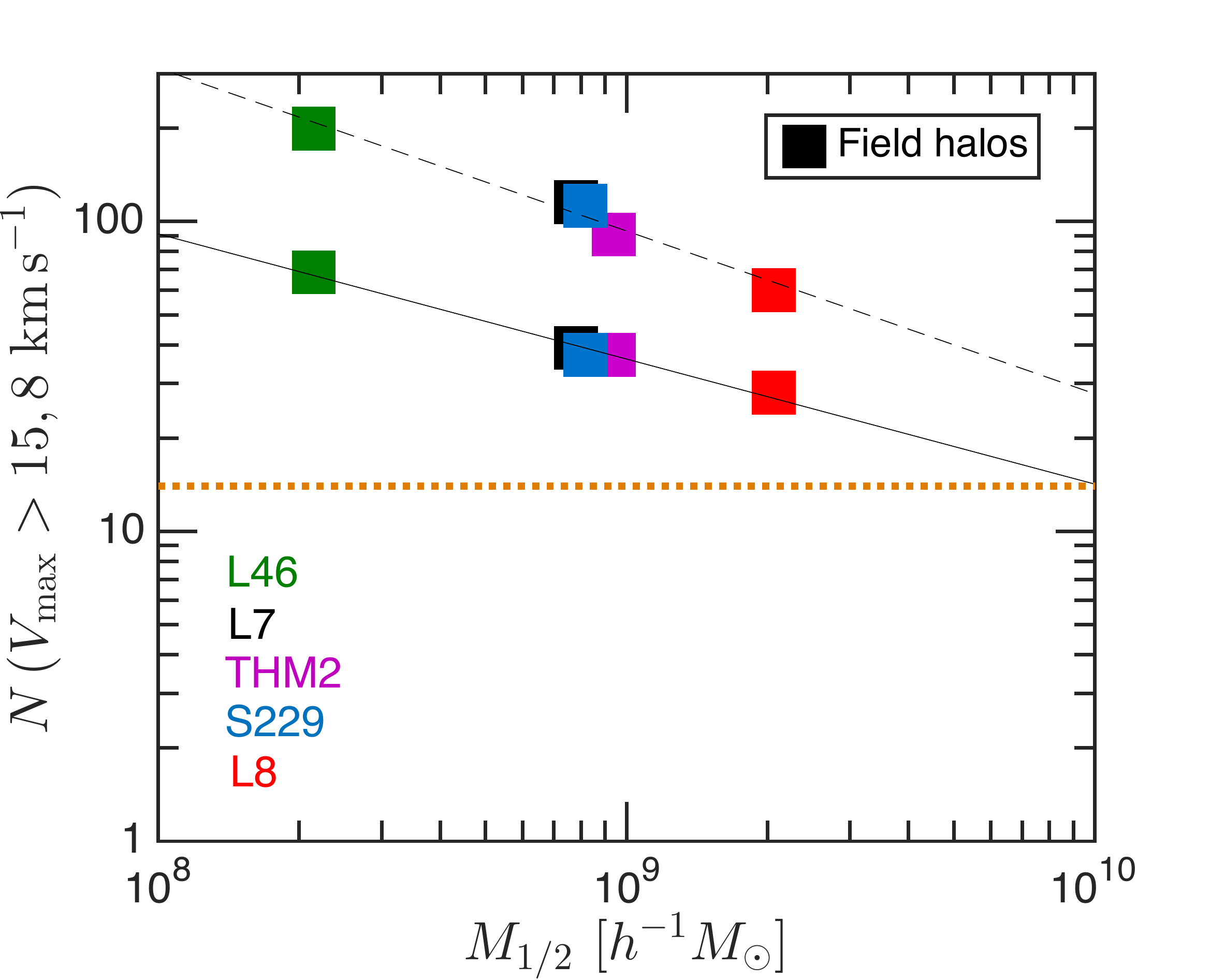}
 \caption{The cumulative number of subhalos above a cutoff velocity in the
   cumulative subhalo $\vmax$ functions for hosts Louise and Thelma 
   (left panel) and the Local Group field halos (right panel) with respect to
   the half-mode mass (defined in the text and listed in Table \ref{tab:modellist}). In the left panel, the solid
   squares (opened diamonds) represent Thelma (Louise) subhalo abundances for DM
   models: L7 (black), THM2 (magenta), L4.6 (green), L8 (red), and S229
   (blue). The solid (dashed) black line gives the power law fit for a cut of $\vmax = 15 \, (8) \, \kms$. The number of DM subhalos in each host halo and in the Local
   Group field is well-predicted by the half-mode mass of the relative transfer
   functions. The orange, dotted horizontal lines show the count of MW satellite galaxies (left; \citealt{Horiuchi2015}) and Local Group field galaxies (right; \citealt{McConnachie2012}). The L8 model is in tension with the current count of MW satellite galaxies.}
\label{fig:Nhalfmode_Nsubs}
\end{center}
\end{figure*}

The general properties of the Thelma and Louise host halos at redshift zero
(e.g., the viral mass, virial radius, physical separation) are similar for each
DM model considered here. The values for these halo properties (among others) can be found in Table \ref{tab:hostprop} for each DM model. The general make-up of the host halos are unaffected by the relative warmth of the respective WDM models. 

The spherically-averaged density profiles and ratios of the WDM to CDM density
profiles for the Thelma and Louise host halos at $z=0$ are shown in Figure
\ref{fig:host_density}. The density profiles for each host show good agreement
for all DM models at all radii. The ratios of the WDM and CDM hosts' density
profiles show only small departures from unity on most scales. The larger
variations near the virial radius in the Louise hosts' DM density profiles is
due to variations in the radial distribution of large substructures. 
Inside the resolution limit of $r = 1\,\hkpc$, the density ratios
show strong fluctuations that are likely the result of numerical effects.

We have also investigated the shape of both host halos in the CDM, L7, THM2, and 
S229 models by comparing the ratios of the minor and intermediate halo axes relative
to the major halo axis. The host halos shapes are triaxial at all radii with some variation in shape with radius. 
The DM model type does not affect the host halo shape as there are no significant 
differences in the Thelma and Louise axis ratios at any radius between CDM and WDM models.

The mass assembly histories of the host halos for each DM model are shown in
Figure \ref{fig:host_assem}. After a redshift of $z = 5$, both host halo assembly histories 
are identical for all DM models. Prior to $z = 5$, the hosts' assembly histories diverge for
each DM model. Although the WDM model hosts collapse later than in the CDM model, there 
is not a correlation between the WDM model warmth and the initial collapse time nor with the subsequent mass build-up 
leading up to the common assembly history of $z < 5$. The host halo mass for each WDM model at the time of 
initial collapse is near or below the half-mode mass, $M_{1/2}$, (as listed in Table \ref{tab:modellist}) where free-streaming effects that suppress structure formation are significant. The early mass assembly of the host halos are then likely built up from the merging of halos with masses on the order of $M_{1/2}$. The model-dependent nature of this mass scale then accounts for the differences in the initial collapse times and early mass assembly history of the different DM model host halos shown in Figure \ref{fig:host_assem}.

\subsection{Local Group Halo Abundances}

The cumulative $\vmax$ functions for Thelma and Louise subhalos, Local Group field 
halos (located within  $1.2$~Mpc of either host center, but outside both host virial radii), 
and a Local Volume set (all DM halos within $1.2$~Mpc of either host center) are
shown in Figure \ref{fig:host_vmaxfunc}. The CDM model predicts hundreds of subhalos 
within each host's virial radius and thousands in the Local Group down to the resolution 
limit of $\vmax \geq 8 \, \kms$. The Thelma, Louise, Local
Group field, and Local Volume cumulative $\vmax$ functions for the WDM models are reduced relative to
the CDM cumulative $\vmax$ function at all masses to a degree that depends on
the warmth of the model; warmer models have a lesser number of subhalos
relative to CDM. The relative suppression in the WDM cumulative $\vmax$
functions is more pronounced at the low-mass end where the strong suppression in
the small-scale power of the transfer functions in Figure \ref{fig:trans_func}
results in greatly reduced low-mass halo formation. For example, the L7 model 
cumulative $\vmax$ functions are reduced by a factor of ten relative to CDM at
the resolution limit, even without removing artificial subhalos.

The cumulative subhalo $\vmax$ functions of both host halos show a strong
similarity between L7 and S229 sterile neutrino models and their thermal
equivalent $m = 2$~keV WDM model over the full $\vmax$ range. There are small differences, on the order of $10\%-20\%$, between the sterile neutrino and thermal equivalent models' low-mass subhalo abundances near the resolution limit. The Local Group and Local Volume $\vmax$ functions also find a slightly greater number of low-mass DM halos ($\sim 20\%$) for the sterile neutrino models than their thermal equivalent model. We caution, however, that these differences are not only near the resolution limit of the simulations, but also set in for $\vmax < 15 \kms$, where artificial fragmentation may remain an issue. The relevance of this difference in the low-mass halo abundance as it relates to satellite galaxy counts in the MW and M31 are discussed in detail by \citet{Horiuchi2015}.

\begin{figure*}
\begin{center}
\begin{tabular}{c c }
\includegraphics[width=0.49\textwidth]{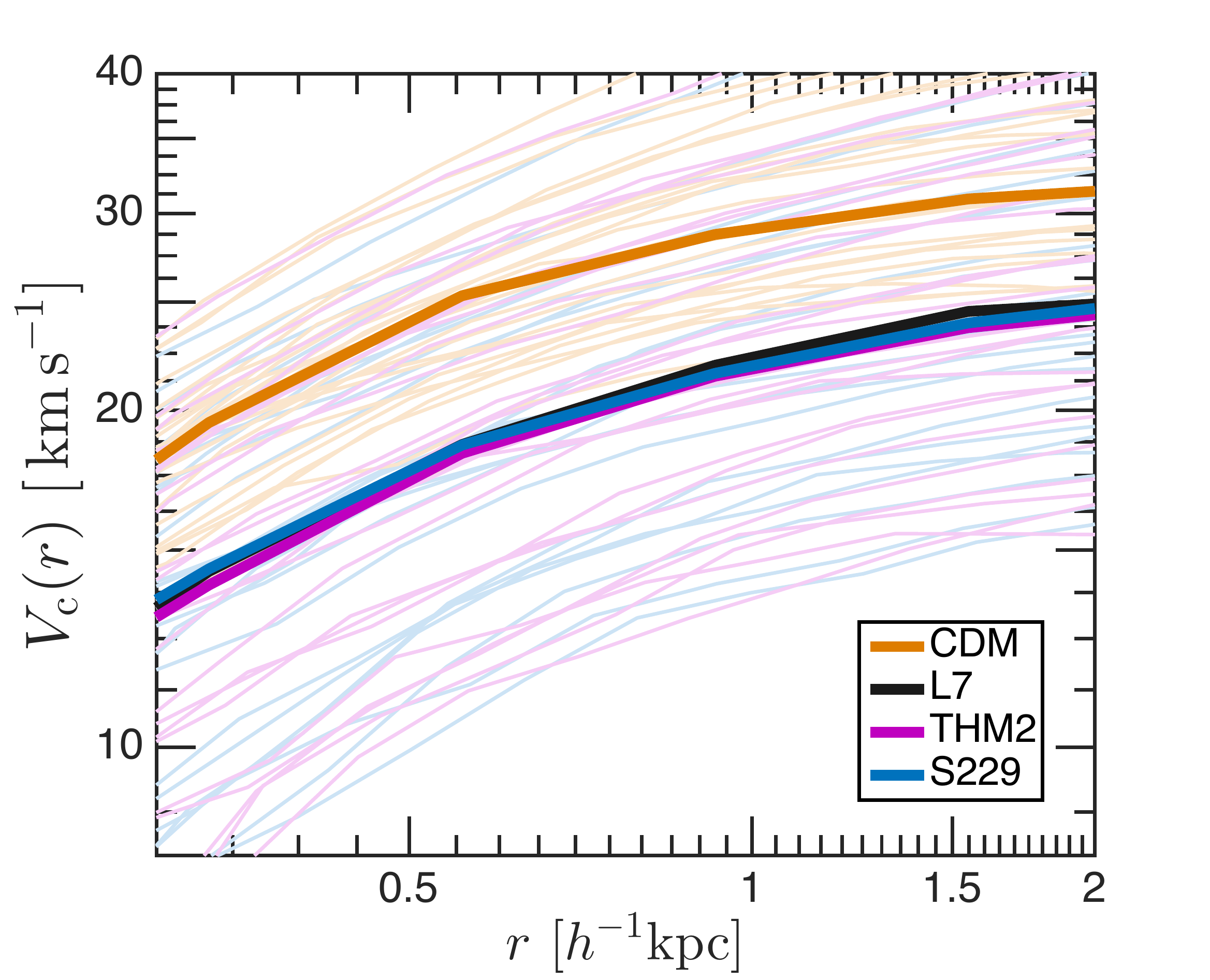} &
\includegraphics[width=0.49\textwidth]{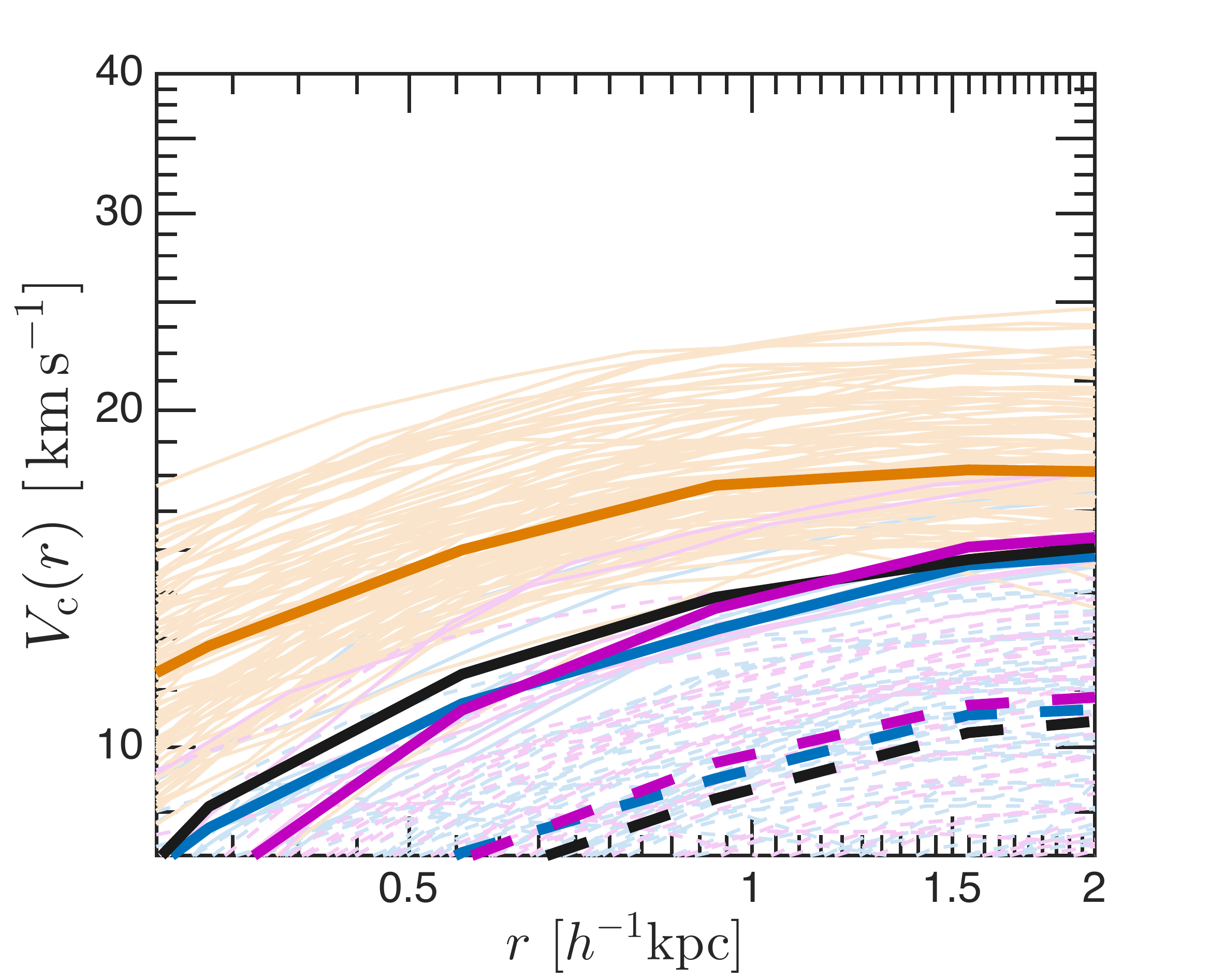} \\
\includegraphics[width=0.49\textwidth]{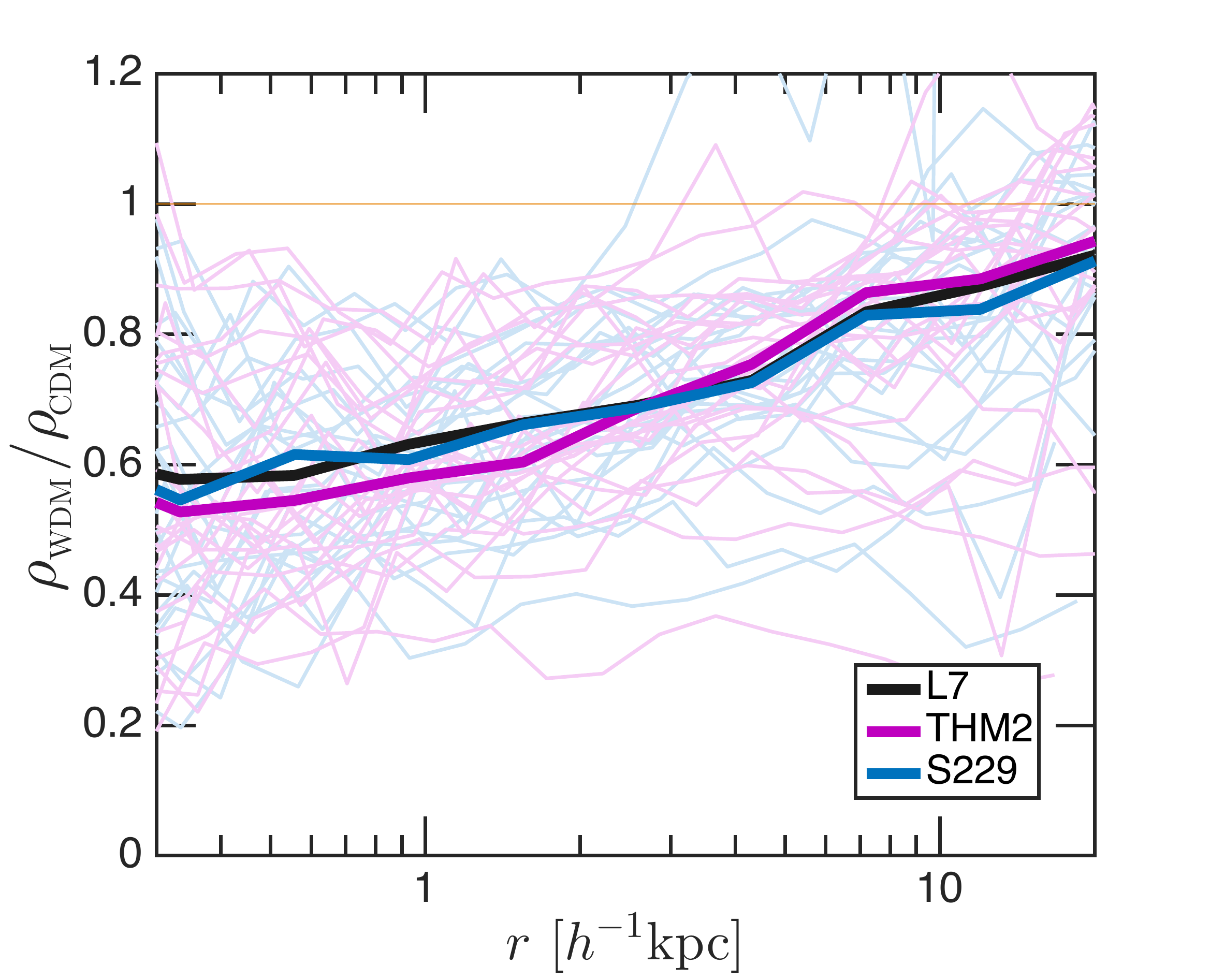} &
\includegraphics[width=0.49\textwidth]{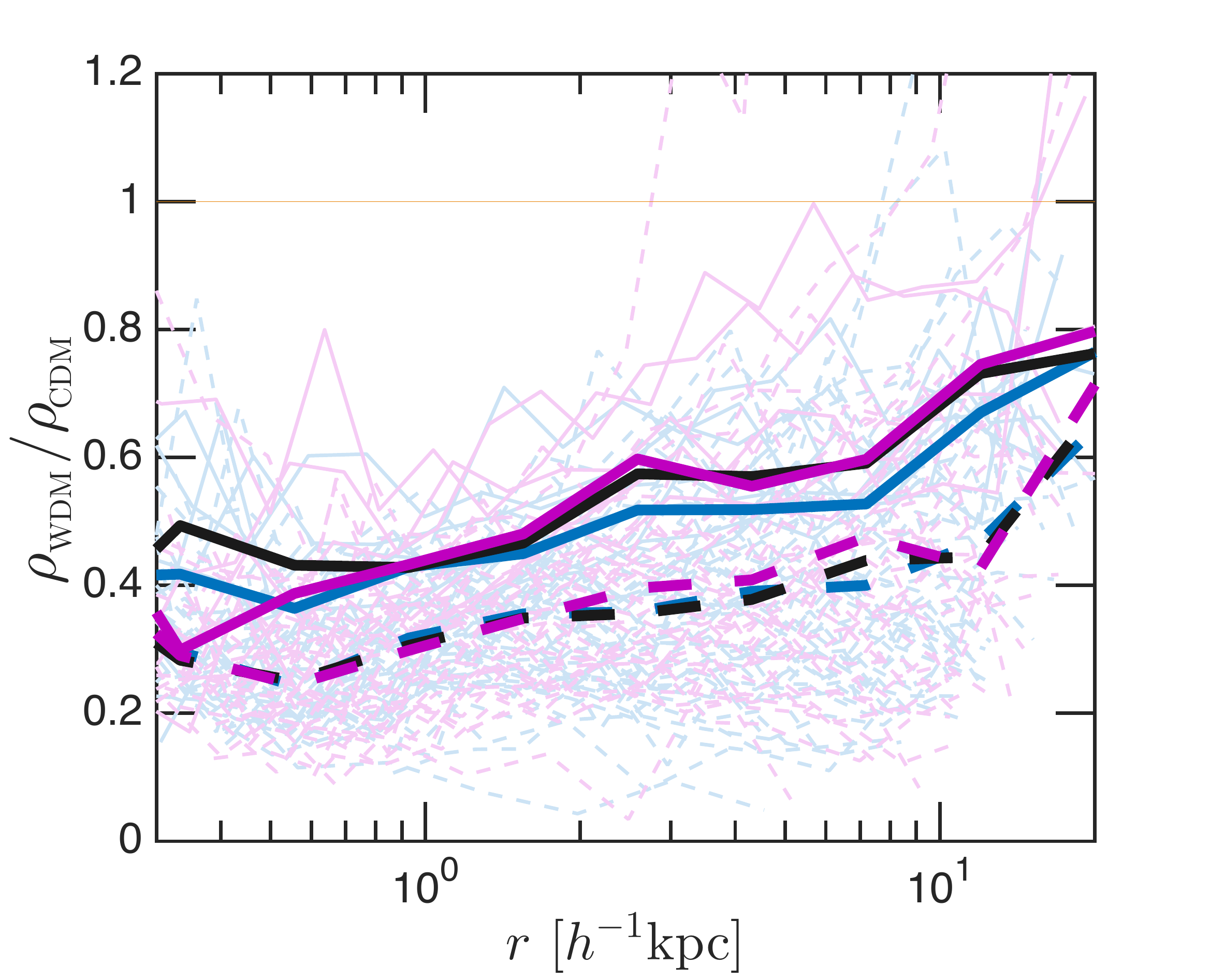} \\
\end{tabular}
\caption{The rotation curves (top row) and density ratios (bottom row) of the
  Local Group field halos for the CDM (orange), S229 (blue), and THM2 (magenta) DM
  models. WDM field halos with $\vmax > 15 \, \kms$ are 
  shown as solid curves and those with $8 < \vmax < 15 \, \kms$ are dashed. The thick dark curves give the median values of each DM model halo set: solid for halos with $\vmax > 15 \, \kms$ and dashed for $\vmax > 8 \, \kms$. The median relations of L7 halos are shown in black (individual L7 halo curves are not shown for clarity). The WDM field halos show a $40\%$ reduction in the median central density relative to CDM halos with $\vmax > 25 \, \kms$ (bottom left panel), and a corresponding ${\sim}5 \, \kms$
  reduction in the median rotation curves (top left panel). Lower
  mass WDM halos (right panels) show a more pronounced decrease in the median
  density ratios of $60-70\%$ in the central $1\, \hkpc$ and $20-30\%$ near the
  halos' virial radius (although with significant scatter). Including halos with $8 < \vmax < 15 \, \kms$
  results in a more pronounced reduction in the WDM median density ratio and rotation curve relative to CDM.}
\label{fig:LGsub_densratios_withmedian}
\end{center}
\end{figure*}

\begin{figure*}
\begin{center}
\begin{tabular}{c c }
\includegraphics[width=0.49\textwidth]{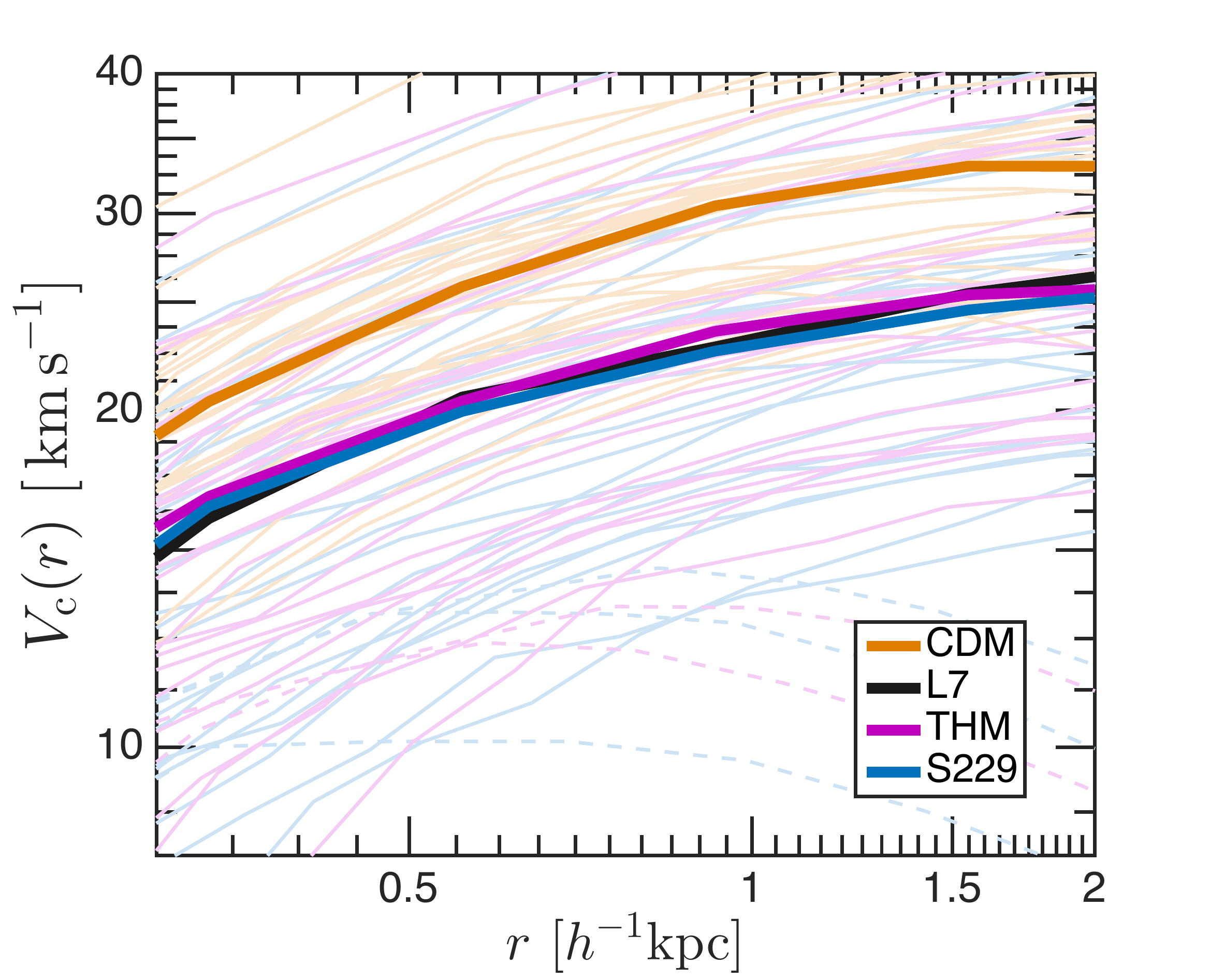} &
\includegraphics[width=0.49\textwidth]{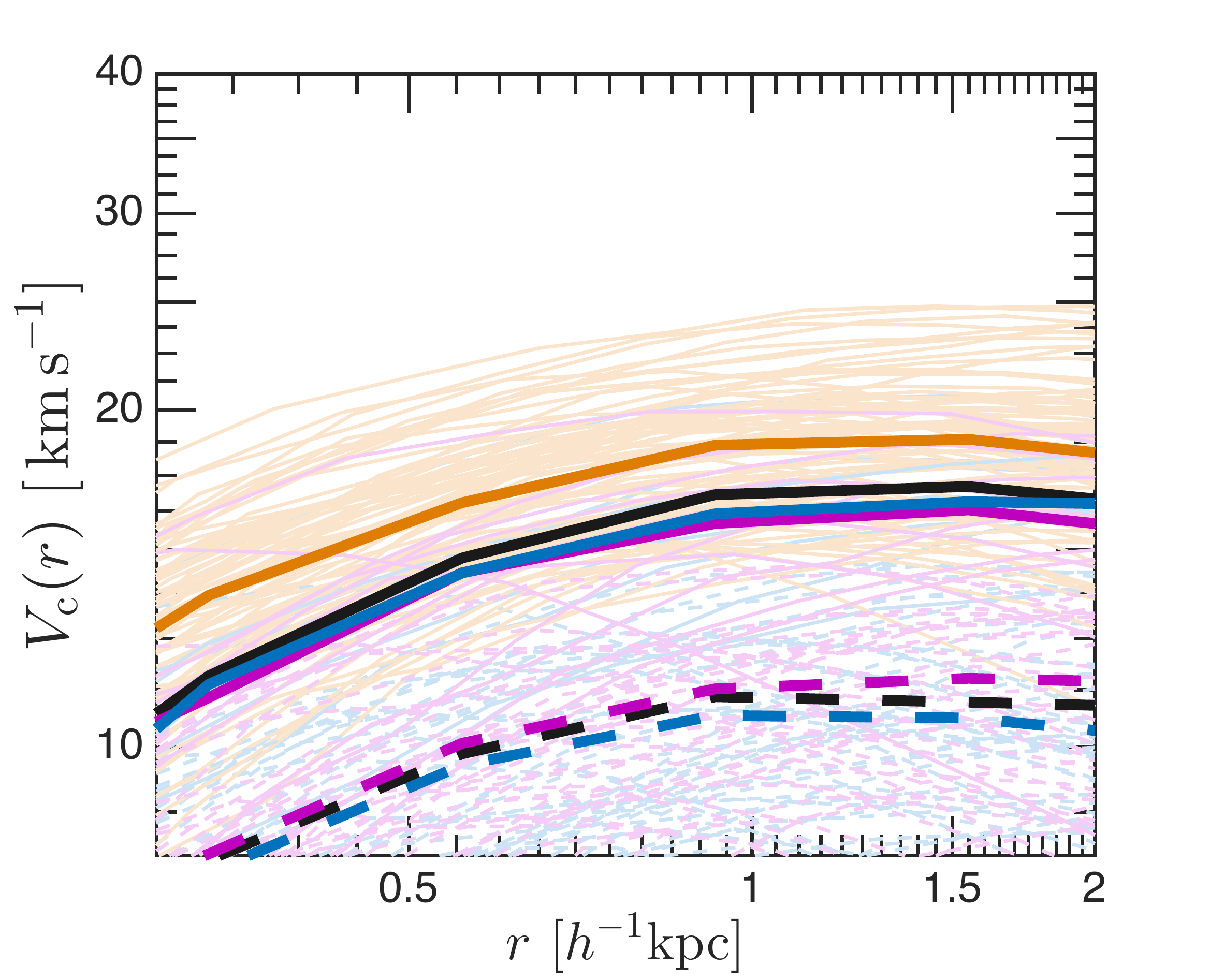} \\
\includegraphics[width=0.49\textwidth]{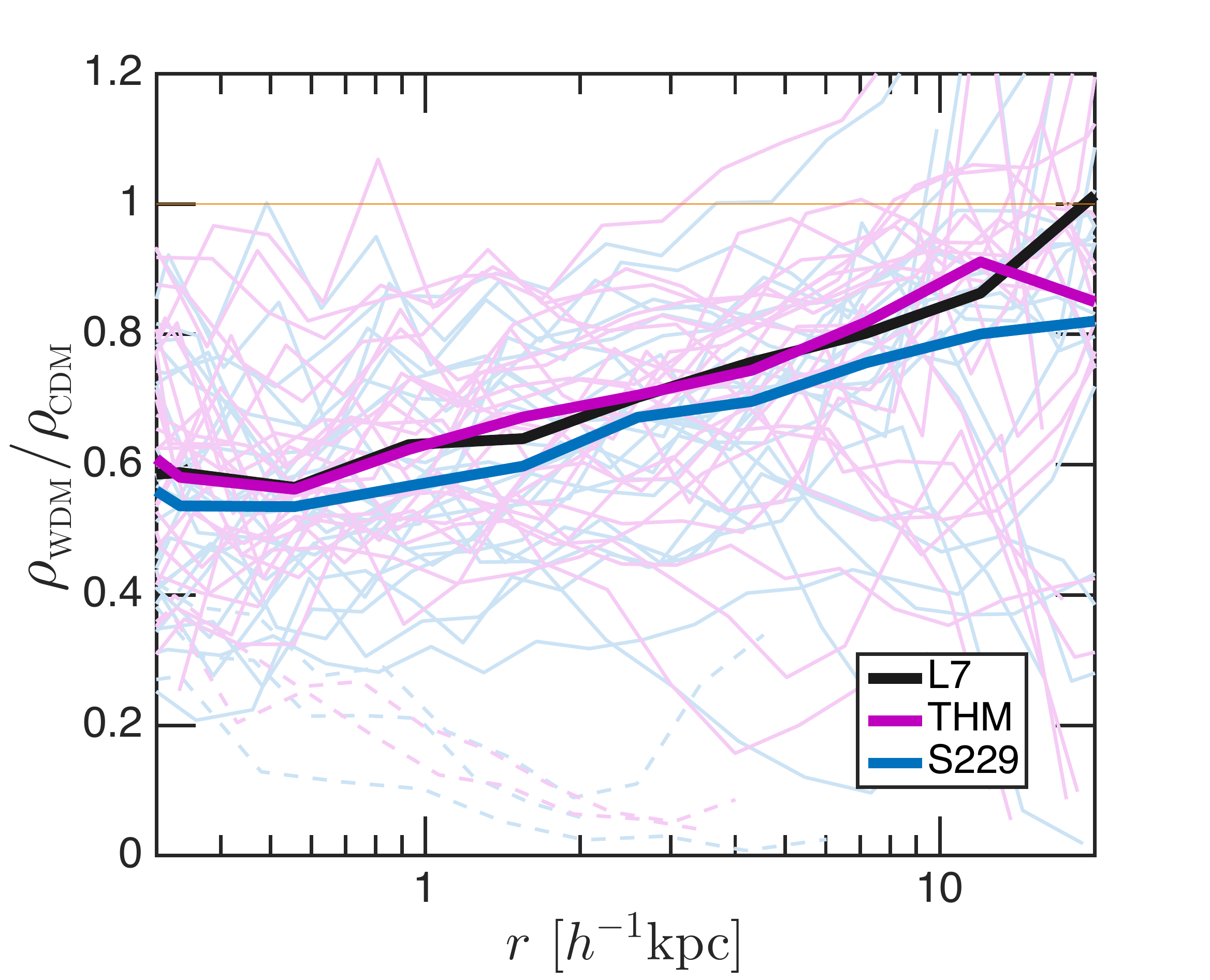} &
\includegraphics[width=0.49\textwidth]{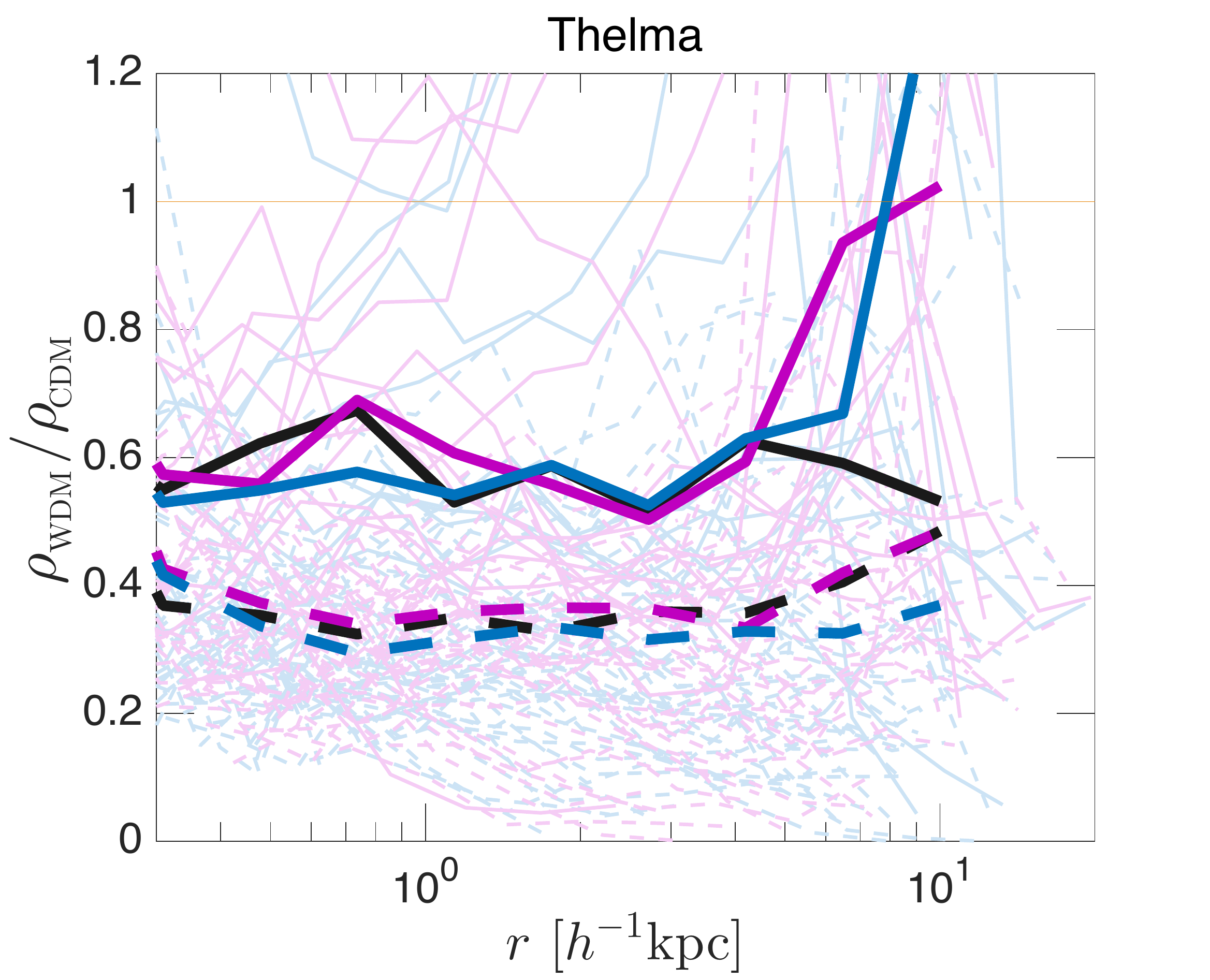} \\
\end{tabular}
\caption{The rotation curves (top row) and density ratios (bottom row) of the
  Thelma and Louise subhalos for the CDM (orange), S229 (blue), and THM2 (magenta) models. The same selection criteria and median curve styles as described Figure \ref{fig:LGsub_densratios_withmedian} are used, with the exception of the bottom right panel, where the median density ratios are determined to $r < 10 \, \hkpc$. The median rotation curves and density ratios of the $\vmax > 25 \, \kms$ subhalos (left column) are quantitatively similar to the Local Group field results shown in Figure \ref{fig:LGsub_densratios_withmedian}. The WDM subhalos counterparts to CDM subhalos with $15 < \vmax < 25 \, \kms$ (right panels) show a $40\%$ reduction in the median central density relative to CDM, if only WDM subhalos with $\vmax > 15 \, \kms$ are considered. There are only $\sim 10$ of these subhalos in each set and there is considerable scatter around the median that is likely due to tidal stripping effects in both WDM and CDM subhalos. Including subhalos with $\vmax > 8 \, \kms$ lowers the median density ratio and rotation curve relative to CDM, bringing them into better agreement with Local Group field halo results.}
\label{fig:TLsub_densratios_withmedian}
\end{center}
\end{figure*}

Figure \ref{fig:Nhalfmode_Nsubs} illustrates the relationship between the
half-mode mass ($M_{1/2}$), the halo mass scale where free-streaming effects are significant, and the total number of halos in the $\vmax$ function above a cutoff velocity. We consider the total number of halos above the cuts of $\vmax \geq 8 \, \kms$ and $\vmax \geq 15 \, \kms$. 
We fit the points in each panel of Fig. \ref{fig:Nhalfmode_Nsubs} to a power law
relation, $y \propto x^n$, where $n = -0.40 \, (-0.51)$ for the combined subhalo set of Thelma and Louise for
$\vmax > 15 \, (8) \, \kms$, and $n = -0.40 \, (-0.53)$ for the field halos in the Local
Group. 

The half-mode mass is strongly predictive of the total number of subhalos in 
both hosts combined (left panel) and in the Local Group (right panel) for
 both $\vmax$ cuts. The subhalo mass function dependence on the filtering mass 
 was previously studied in \citealt{Dunstan2011} for non-resonant WDM models. 
 The power law trends of the host subhalo sets and the Local Group field halos are very similar.
 Warmer WDM models show a larger dispersion 
 around the power-law fit for Thelma and Louise subhalo abundances. This
is likely due to the small number of subhalos in the warmer models, since the
variation around the trend is stronger for the larger $\vmax$ cut and the trend
in Local Group for the larger $\vmax$ cut is the same as the subhalo set. The
slight difference in the half-mode mass between thermal and sterile neutrino
models results in a small difference in the total number of halos near the resolution limit 
in each host and in the Local Group.

\begin{figure}
\begin{center}
\includegraphics[width=0.49\textwidth]{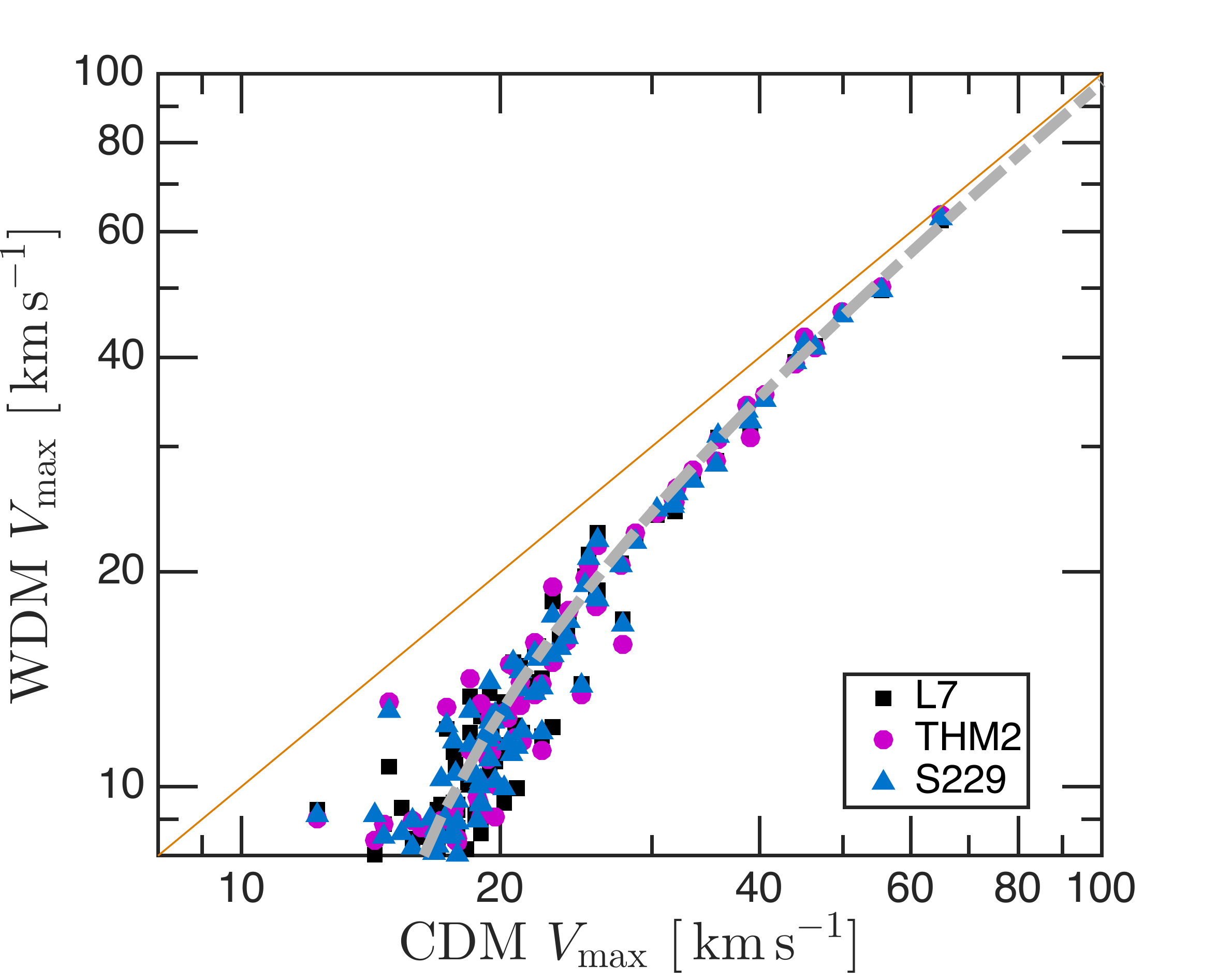}
\caption{The WDM $\vmax$ as a function of the matched CDM $\vmax$ on a halo-by-halo basis for Local Group field halos. The
  different WDM model halos (L7 - black; S229 - blue; THM2 - magenta) have a
  similar distribution that bends away from the a one-to-one relation (orange
  curve) with CDM halos for smaller halo masses, as a result of reduced central
  densities in the WDM halos. The gray curve shows a fit to the points that
  includes all WDM model halos. Fits to individual WDM model halo sets (not shown) are
  similar to the fit for all WDM model halos.}
\label{fig:Vmax_Relation}
\end{center}
\end{figure}

\subsection{Local Group Subhalo and Field Halo Internal Structure}

We next consider the internal structure of the Local Group field halos and Thelma and
Louise subhalos. The rotation curves and density profile ratios of WDM relative
to CDM halos for the S229, L7, and THM2 Local Group field halos are shown in Figure
\ref{fig:LGsub_densratios_withmedian}. We also determine the median rotation curves and median 
density profile ratios for each DM model halo set (thick, dark curves). 
For simplicity, the individual L7 halo curves are not shown 
(although the median values of the L7 sets are presented). 
The halo sets were selected by, first, identifying all CDM Local Group field halos (located within  $1.2$~Mpc of either host center, but outside both host virial radii). CDM Local Group field halos are then divided into two mass bins: 
halos with $\vmax > 25 \, \kms$ (left column panels) and $15 < \vmax < 25 \, \kms$ 
(right column panels). We then identify WDM Local Group field halo 
counterparts to these CDM halos that satisfy the halo correlation criterion described in Section
\ref{sec:Sims}. 

The WDM halo counterparts are required to also be located 
in the Local Group field, but not to follow the $\vmax$ cuts given above, i.e. a WDM halo counterpart could 
be significantly more or less massive than the corresponding CDM halo. The THM2 and S229 WDM model 
rotation and density ratio curves, shown in Figure \ref{fig:LGsub_densratios_withmedian}, are solid 
for halos with $\vmax > 15 \, \kms$ and dashed for halos with $8 < \vmax < 15 \, \kms$. The CDM field halos with $\vmax > 25 \, \kms$ (left column) 
have all WDM halo counterparts with $\vmax > 15 \, \kms$. For CDM field halos with $15 < \vmax < 25 \, \kms$ (right column), however, 
only a handful of WDM halo counterparts have a $\vmax > 15 \, \kms$; most have a $\vmax$ value of  
$8 < \vmax < 15 \, \kms$. Since the majority of WDM counterparts to the CDM field halos with $15 < \vmax < 25 \, \kms$ have a $\vmax$ value 
below the artificial fragmentation limit, we plot the median rotation curve and median density profile ratio for WDM counterparts with $\vmax > 15 \, \kms$ (thick, solid curves) and for WDM halos with $\vmax > 8 \, \kms$ (thick, dashed curves) in the right column panels of Figure \ref{fig:LGsub_densratios_withmedian}. While the WDM halo sets with $\vmax > 8 \, \kms$ are possibly contaminated by artificial fragmentation, they represent the majority of WDM counterparts to the CDM field halos in the low-mass bin. If artificial fragmentation is not 
significant, then the WDM field halos with $\vmax > 8 \, \kms$ are the appropriate comparisons to CDM field halos with $15 < \vmax < 25 \, \kms$. There is also a significant fraction of WDM counterpart halos to CDM field halos with $15 < \vmax < 25 \, \kms$ that have a $\vmax$ below our resolution limit, $\vmax < 8 \, \kms$ (not shown). We do include these halos in the subsequent analysis.

The median rotation curves of the L7, THM2 and S229 WDM models, shown in the top 
row panels of Figure \ref{fig:LGsub_densratios_withmedian}, fall uniformly below
the CDM rotation curves for both $\vmax$ sets. The reduction in the median WDM
rotation curves relative to CDM is more pronounced in the lower mass halo sets
(top right panel) for the WDM halos with $\vmax > 8 \, \kms$. 
The median rotation curves of the L7, S229, and THM2 WDM models are similar for both
$\vmax$ sets, indicating that each WDM model has, on average, a similar reduction
in the central density of their DM halo relative to CDM.

The bottom panels of Figure \ref{fig:LGsub_densratios_withmedian} show the ratio
of the WDM field halo density profiles to CDM. The median density ratios of the Local Group WDM
field halo counterparts to CDM field halos with $\vmax > 25 \, \kms$ (bottom left panel) show that the CDM and WDM
density profiles converge in the outer halo near $r = 20\,\hkpc$. The inner region of
WDM density profiles reach an average reduction of ${\sim}40\%$ at
$r \leq 1\,\hkpc$. For CDM field halos with $15 < \vmax < 25 \, \kms$, the median density ratios of
 the corresponding Local Group WDM field halo sets (bottom right panel) show the WDM halos 
with $\vmax > 15 \, \kms$ are less dense than their CDM counterparts throughout, and are on average ${\sim}60\%$
less dense at $r = 1\,\hkpc$. Including the WDM field halos with $\vmax > 8 \, \kms$ in the comparison lowers 
the median density ratio an additional $10\%$ in the central $r \leq 1\,\hkpc$ and an additional $20\%$ at $r = 10\,\hkpc$ 
near the average halo virial radius. The median density ratios of the L7, S229, and THM2 WDM models are quantitatively 
similar at all radii for both $\vmax$ sets and independent of the limiting WDM halo mass. 

The reduction in the WDM central density is
related to later formation times of WDM halos relative to CDM halos, which
collapse earlier when the background matter density of the universe is greater.
For both $\vmax$ sets, there is considerable scatter in the individual halo density
ratios at all radii. This spread in internal density ratios is reflected in the WDM rotation
curves of the both mass set in the top panels. The CDM rotation curves are
tightly clustered around the median, while the WDM sets show a larger degree of
scatter. We find that, for an individual CDM halo, it is difficult to predict the exact amount of reduction in 
the corresponding WDM halo density profile (and likewise the rotation curve), however, 
in general, the less massive the CDM field halo the greater the reduction in the density profile 
(rotation curve) at all radii. 

The internal properties of the WDM subhalos of the Thelma and Louise host halos
are also significantly different than their CDM counterparts. We follow the same selection process 
for the host subhalos sets, with the exception that the CDM subhalos and their WDM counterparts are required 
to be within the virial radius of either host. For the $\vmax > 25 \, \kms$ subhalo sets, shown in the left column of 
 Figure \ref{fig:TLsub_densratios_withmedian}, the median rotation curves 
 and median density profile ratios of each DM model are quantitatively similar to the median values of the 
 $\vmax > 25 \, \kms$ Local Group field halos. The L7, S229, and THM2 WDM models each have a 
${\sim}40\%$ reduction at $r \leq 1\,\hkpc$ in the median density ratios and a ${\sim}5 \, \kms$ drop 
in the median WDM rotation curves relative to the median CDM rotation curve, 
just as we found for the Local Group field halos. We find no significant difference in the L7, S229, and THM2 model median 
rotation curves or density profile ratios.  

The median rotation curve of CDM subhalos with $15 \, \kms < \vmax < 25 \, \kms$ is also similar to the CDM 
Local Group field halo median rotation curve. The median rotation curves of the WDM subhalo counterparts with 
$\vmax > 15 \, \kms$ (shown in the top right panel of Figure \ref{fig:TLsub_densratios_withmedian}), however, 
have only a few $\kms$ reduction in magnitude relative to CDM. The median density 
ratios of the WDM subhalos with $\vmax > 15 \, \kms$ show only a $40\%$ reduction in the central density relative to CDM. 
This is a smaller decrease than was found for WDM Local Group field halos of the same $\vmax$ cut. There is considerable 
scatter around the median for the individual density profile ratios of the WDM subhalos with $\vmax > 15 \, \kms$ 
that is likely due to tidal stripping effects in both the CDM and WDM subhalos. Including the WDM subhalos with $\vmax > 8 \, \kms$ brings better agreement with the Local Group field halo results for the same $\vmax$ selection criteria. 
The median WDM density profiles are then reduced relative to CDM by $60\%$ at all radii out to $r = 10\,\hkpc$, 
and the median WDM rotation curves show a more significant decrease on the order of ${\sim}10 \, \kms$. 
For the subhalo sets, we also find no significant difference in the L7, S229, and THM2 model subhalo internal properties. 
Similar to the Local Group field halo results, we find that, on average, the less massive the CDM subhalo, the greater 
reduction in the corresponding WDM density profile. Tidal stripping effects, however, can complicate a comparison of a single subhalo between CDM and WDM models. 

Figure \ref{fig:Vmax_Relation} shows the $\vmax$ relation between the Local
Group field halos in WDM and CDM. Halos were selected by choosing Local Group CDM field halos with a
$\vmax > 8 \, \kms$ and then identifying WDM halo counterparts that satisfy the
halo correlation criterion described in Section \ref{sec:Sims}. The orange curve shows the
one-to-one correspondence between WDM and CDM. All Local Group WDM field halos fall
below the orange curve and show a trend of increasingly smaller $\vmax$ relative
to the CDM values. The gray dashed curve in Figure \ref{fig:Vmax_Relation} shows
a fit to the $\vmax$ relation of
\begin{equation}
  \label{eq:vmax_fit}
 \frac{\vmax({\rm WDM})}{\vmax({\rm CDM})}=1-\left(\frac{\vmax({\rm CDM})}{V_{0}}\right)^{\beta}\, ,
\end{equation}
where $V_{0} = 10.6 \, \kms$ and $\beta = -1.58$ for all WDM halos combined. The individual fits to the L7, S229,
and THM2 models are not shown, but each is very similar to the combined fit. We note that $\vmax$ values below $\vmax = 15 \, \kms$ were included in order to obtain a more accurate accounting
of the turnover in the $V_{\rm max,WDM} - V_{\rm max,CDM}$ relation. The $V_{\rm max,WDM} - V_{\rm max,CDM}$ relation summarizes the increased suppression of the WDM halo central density relative to CDM at smaller halo masses.

\section{Abundance of Field Halos}
\label{sec:FieldHaloAbund}

\begin{figure}
\begin{center}
\includegraphics[width=0.49\textwidth]{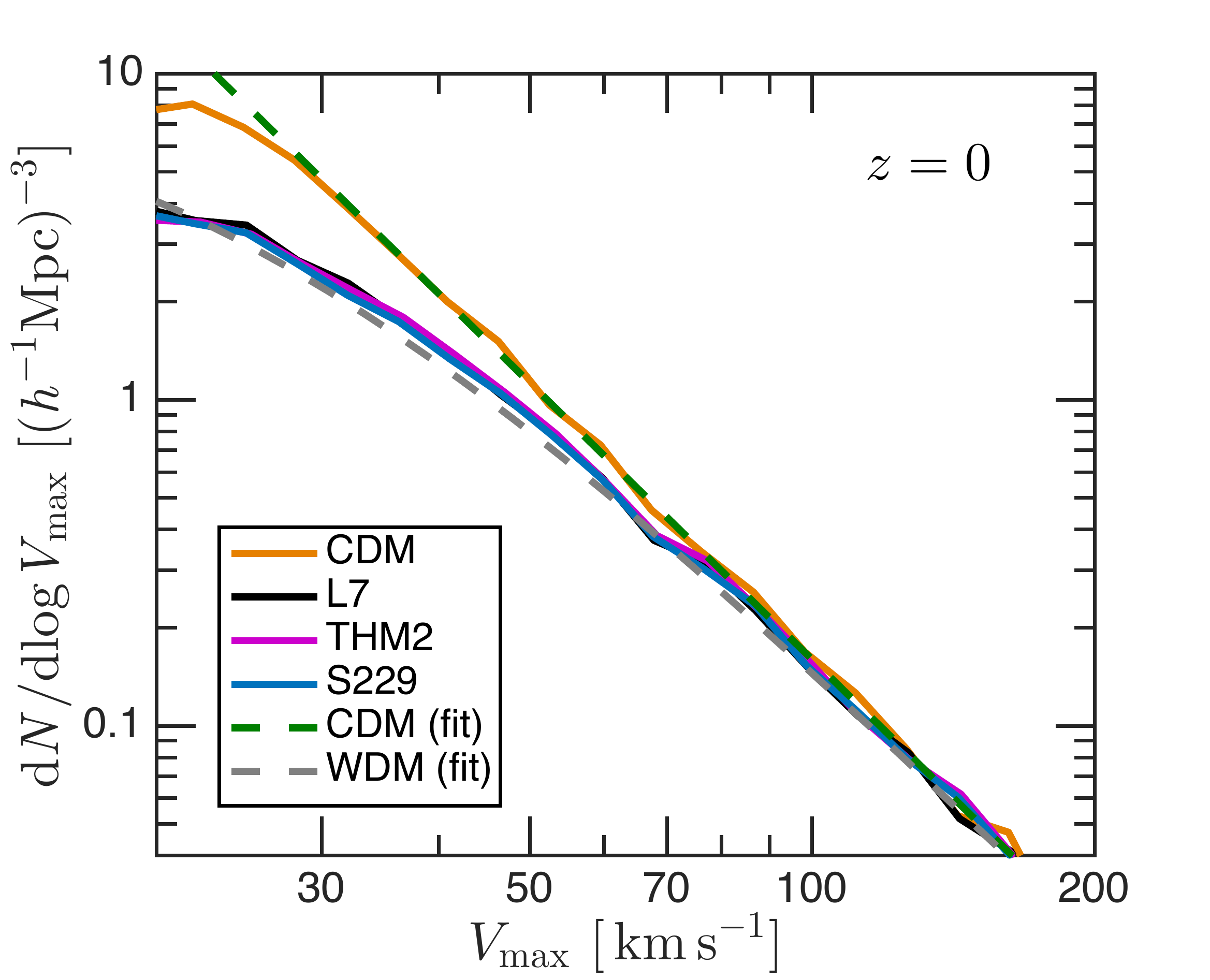}
\caption{The circular velocity functions of field halos in cosmological 
volume simulations for the CDM (orange), L7 (black), S229 (blue) and
  THM2 (magenta) DM models at redshift $z = 0$. The dashed green and grey curves respectively
  represent the CDM and thermal $m = 2$~keV WDM analytic approximations to the
  velocity functions. Our results show a good
  agreement with the analytic fits down to $V = 30 \, \kms$. 
  The sterile neutrino and thermal WDM velocity functions
  are similar for the full range of halo masses and have a maximum suppression 
  of a factor of 2 relative to low-mass CDM halo abundances. }
\label{fig:fullbox_vmaxfunc_z0}
\end{center}
\end{figure}

\begin{figure}
\begin{center}
\includegraphics[width=0.49\textwidth]{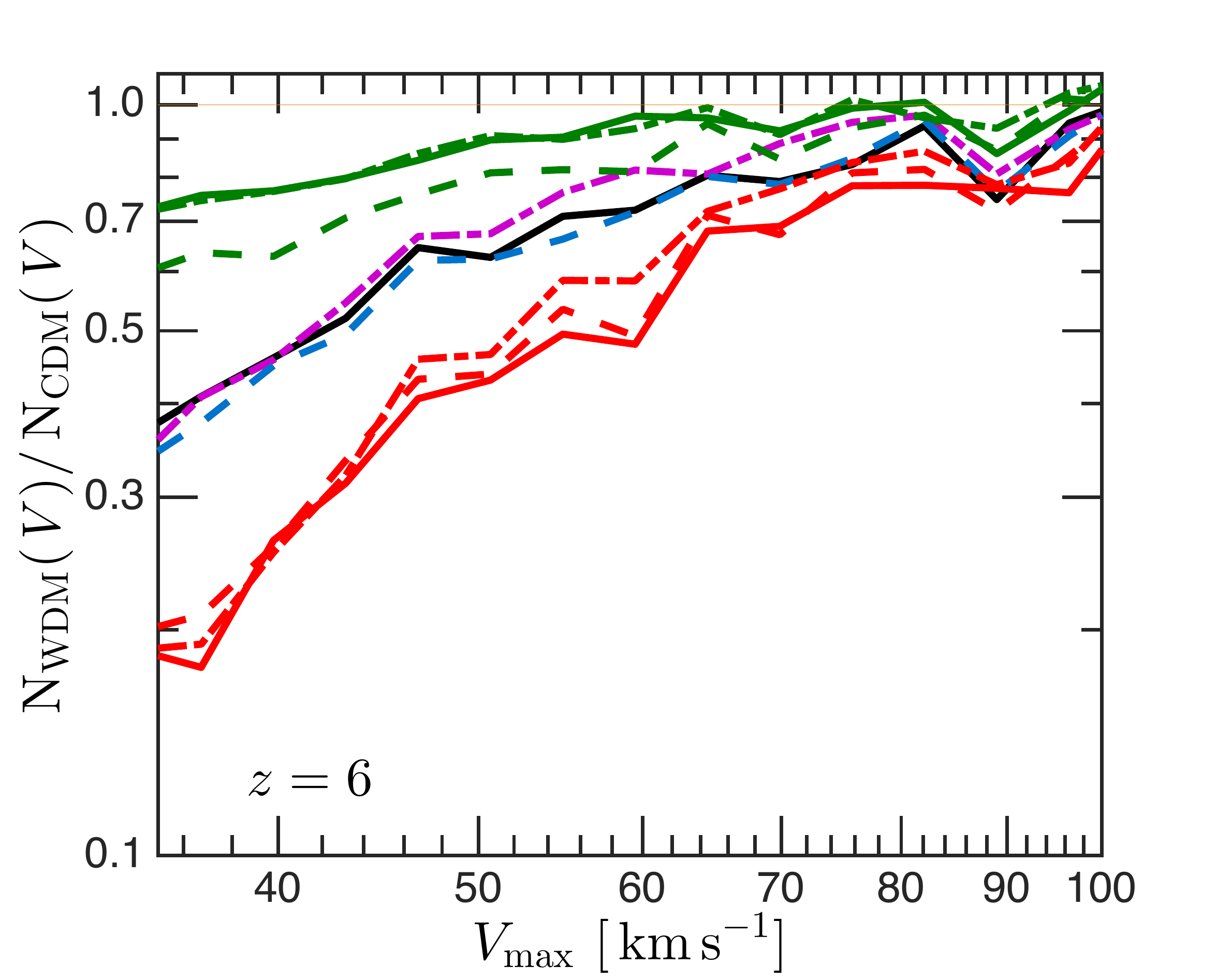}
\caption{The ratio of the circular velocity functions of WDM models to CDM at a 
redshift of $z =6$. The WDM models shown: L46 (solid; green), S220 (dashed;
  green), THM2.9 (dash-dot; green), L7 (solid; black), S229 (dashed; blue), THM2
  (dash-dot; magenta), L48 (solid; red), S208 (dashed; red), THM1.6 (dash-dot;
  red). The WDM model velocity function ratios cluster into three groups according to
  their respective warmth, with only a small ${\sim}5\%$ variation in the velocity 
  function ratios between WDM models of comparable warmth.}
\label{fig:fullbox_vmaxfunc_6}
\end{center}
\end{figure}

The Missing Satellites and TBTF problems of CDM in the Local Group persist in the abundance
and kinematics of field dwarf galaxies \citep{Zavala2009,
  Klypin2015,Papastergis2015} motivating the exploration of nonlinear
evolution of WDM models in cosmological volumes. The redshift zero velocity function of field DM halos
in a homogeneously-resolved, $N_p = 512^3$ {\it N}-body simulation of a
$L = 25\,\hmpc$ box for the L7, S229, THM2 and CDM models are shown in Figure \ref{fig:fullbox_vmaxfunc_z0}. 
The CDM velocity function increases monotonically towards small mass down to $\vmax = 30 \, \kms$. 
We calculate a power law fit to the CDM velocity function that is based on the analytic approximation of 
\citet{Klypin2015} to the Bolshoi simulation results \citep{Klypin2011}. Our fit is very similar to the \citet{Klypin2015} fit 
with a power law exponent of $n = -2.8$ and a $\sim20\%$ increase in the normalization. The analytic fit is shown in 
Figure \ref{fig:fullbox_vmaxfunc_z0} to be in good agreement with the simulation results down to $\vmax = 30 \, \kms$, 
below which numerical effects cause the CDM velocity function to turn over. 

The L7, S229, and THM2 WDM model
velocity functions agree with the CDM velocity function for larger mass halos
with $\vmax > 80 \, \kms$, but are suppressed relative to CDM for low-mass halos,
reaching a factor of two suppression at $\vmax = 30 \, \kms$. The sterile neutrino models and
thermal WDM velocity functions agree with each other and an analytic fit to a $m = 2$~keV thermal WDM 
velocity function, where the equation is taken from \cite{Klypin2015} and modified by our CDM power law exponent and 
normalization values. The low-mass suppression in the WDM velocity function relative to CDM is related to the reduction in 
the WDM halo central density profiles found in the previous section and shown in Figure \ref{fig:Vmax_Relation}. For halos of the mass range considered here, the WDM halos at a fixed abundance are the counterparts to CDM halos of the same abundance, which corresponds to a leftward shift of the CDM velocity function to the WDM velocity function. However, at smaller masses below the resolution of these simulations, the suppression of structure formation in WDM models would become more prevalent, resulting in CDM halos where a WDM counterpart does not form. This would correspond to a decrease in WDM halo abundance relative to CDM at fixed $\vmax$. 

\begin{figure*}
\begin{center}
\begin{tabular}{c c }
\includegraphics[trim=0.5mm 0.5mm 0.5mm 0.5mm,clip,width=0.49\textwidth]{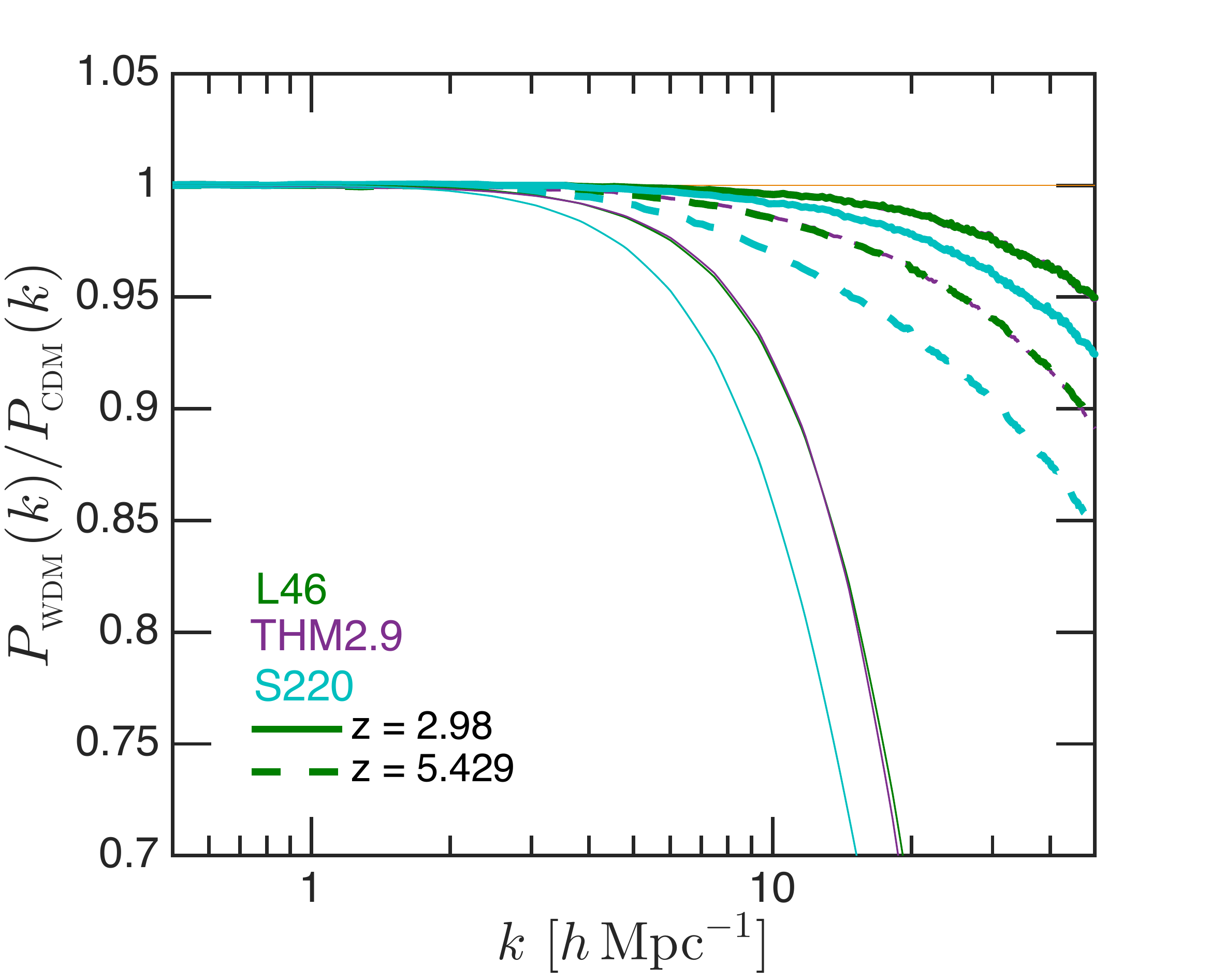} &
\includegraphics[trim=0.5mm 0.5mm 0.5mm 0.5mm,clip,width=0.49\textwidth]{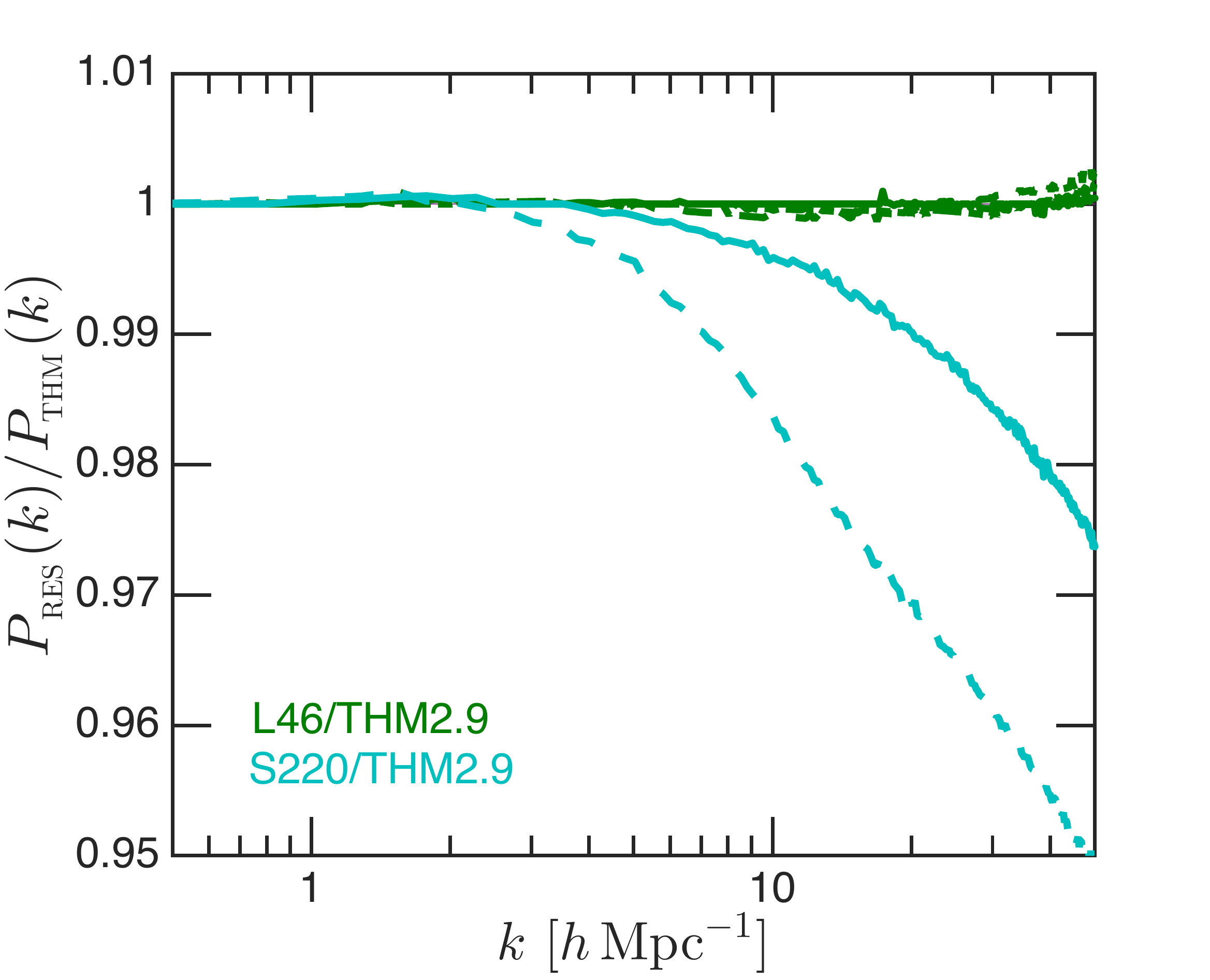} \\
\includegraphics[trim=0.5mm 0.5mm 0.5mm 0.5mm,clip,width=0.49\textwidth]{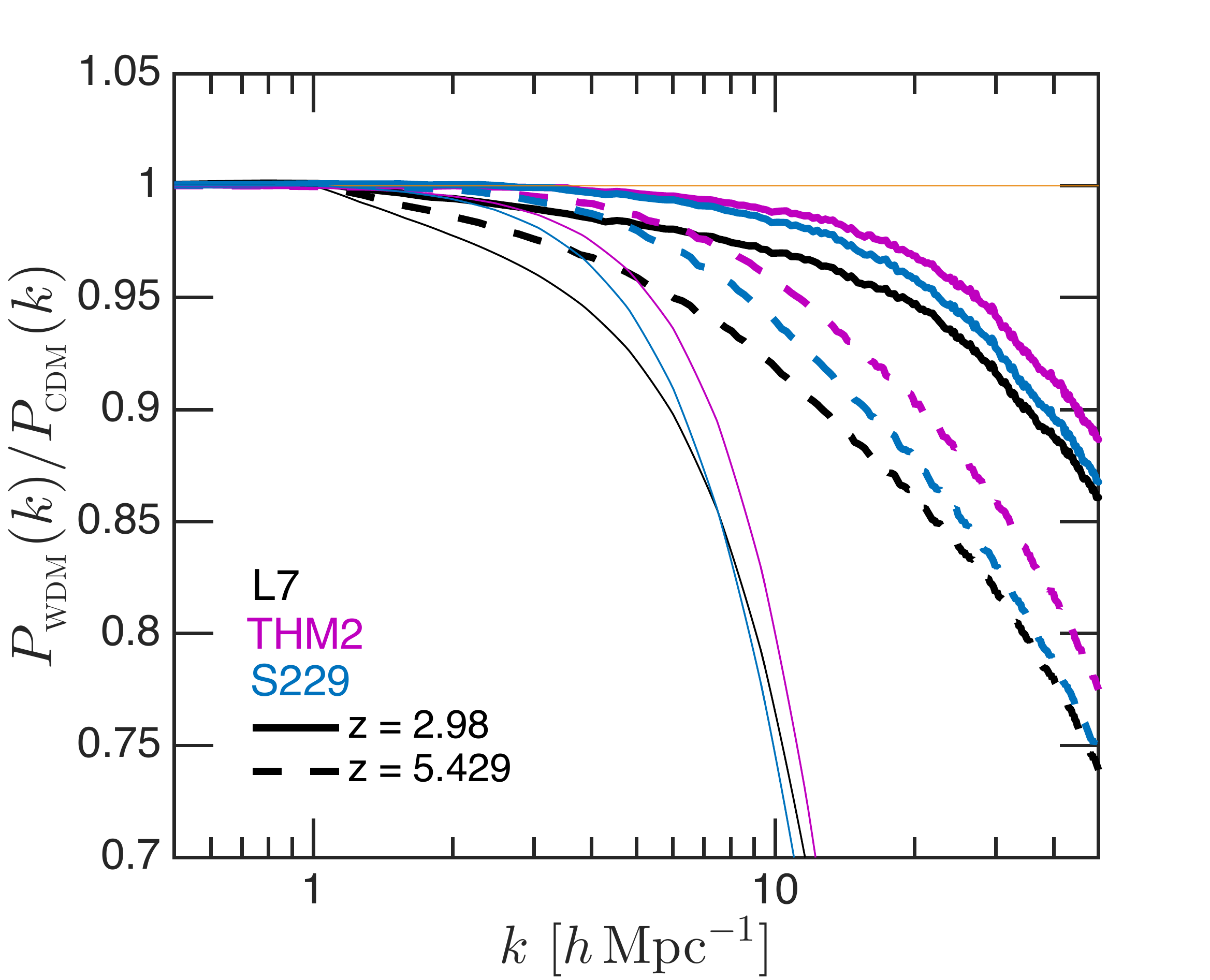} &
\includegraphics[trim=0.5mm 0.5mm 0.5mm 0.5mm,clip,width=0.49\textwidth]{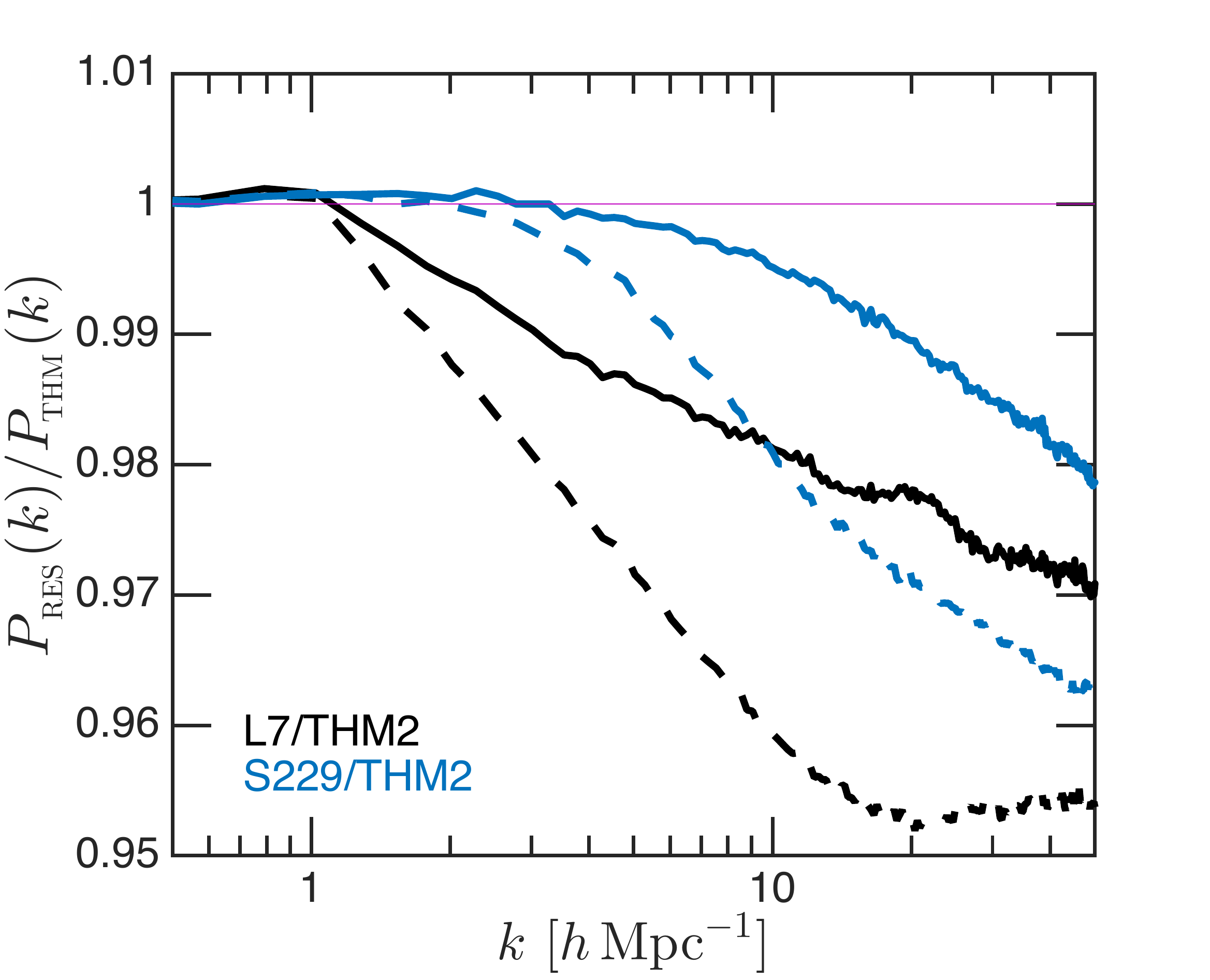} \\
\includegraphics[trim=0.5mm 0.5mm 0.5mm 0.5mm,clip,width=0.49\textwidth]{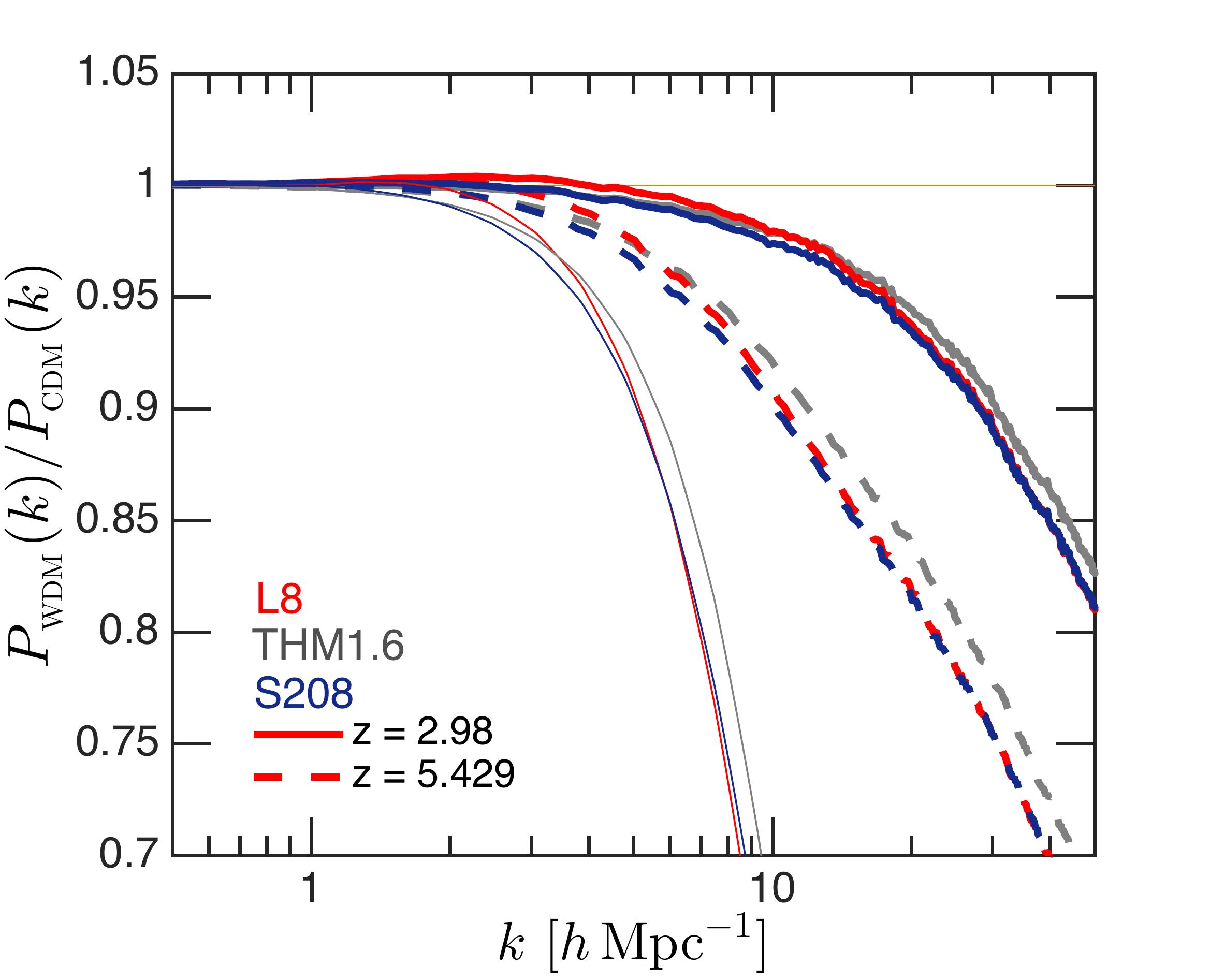} &
\includegraphics[trim=0.5mm 0.5mm 0.5mm 0.5mm,clip,width=0.49\textwidth]{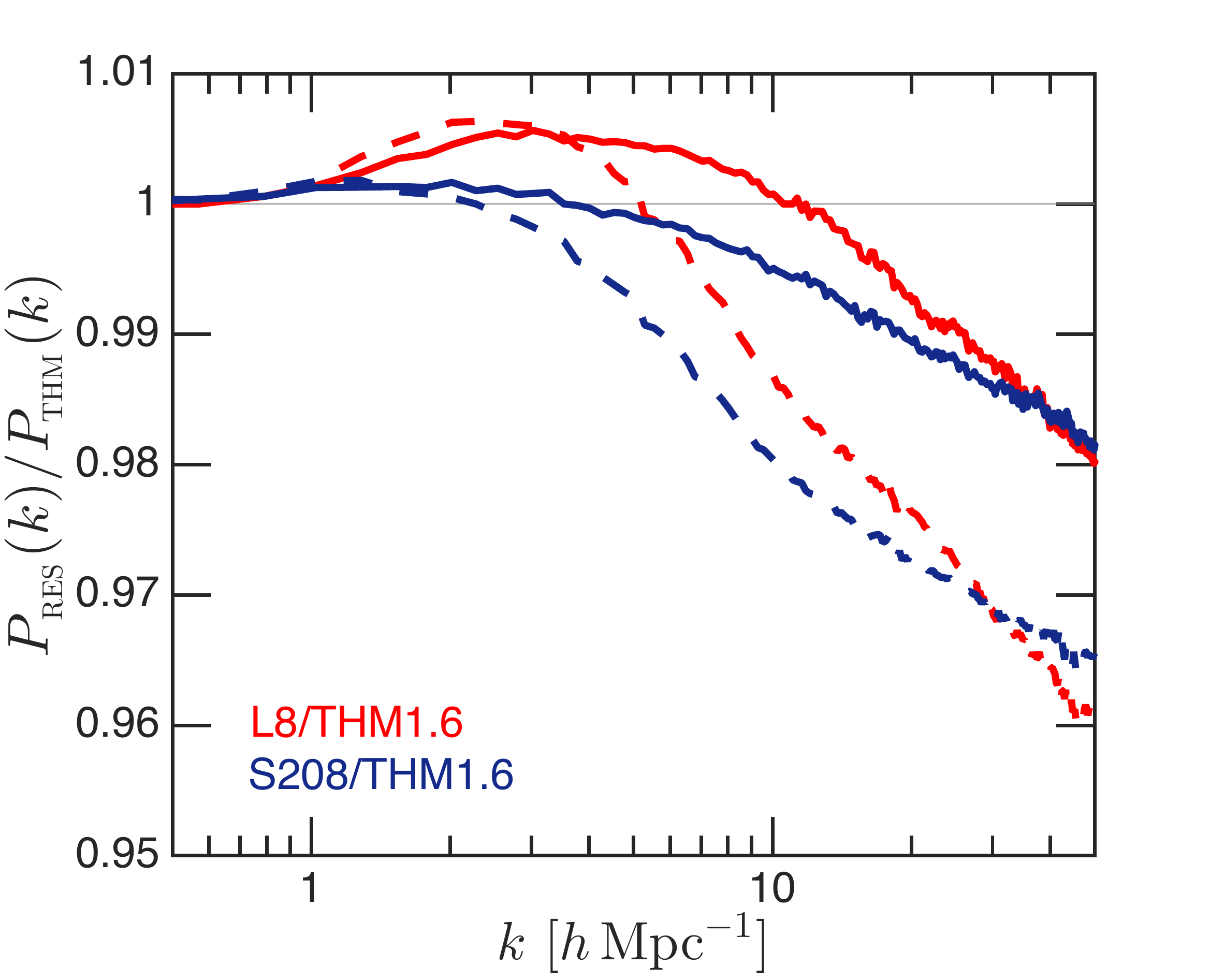} \\
\end{tabular}
\caption{{\it Left column:} ratio of the 3D matter power spectrum for WDM models with respect
  to CDM for redshifts $z = 2.98$ (solid) and $z = 5.429$ (dashed). 
  {\it Right column:} ratio of the 3D matter power spectrum of sterile neutrino models with respect
  to a thermal equivalent WDM model. Rows divide WDM models into groups of similar model warmth. 
  {\it Top Row:} L46 (green), S220 (cyan), THM2.9 (purple). {\it Middle Row:} L7 (black), S229 (blue), and THM2 (magenta). 
  {\it Bottom Row:} L8 (red), S208 (blue), THM1.6 (grey). Solid, thin curves (left column) are the linear matter power spectrum ratios of WDM to CDM. The 3D matter power spectrum of the sterile neutrino and thermal WDM models
  suppress small-scale power relative to CDM, but to differing degrees depending
  on the warmth of the model. The matter power spectrum ratios on scales above
  $k > 1\, \hmpcinv$ show differences at the few percent level, which is significant, between
  the sterile neutrino and thermal WDM models.}
\label{fig:fullbox_ps}
\end{center}
\end{figure*}

The ratio of the velocity function between WDM and CDM models at redshift
$z = 6$ are shown in Figure \ref{fig:fullbox_vmaxfunc_6}. We consider a broader
range of models here. The velocity function ratios of all the A14 (solid curves) and V15 (dashed curves) 
sterile neutrino models and their thermal equivalent models (dot-dashed curves) are shown. The relative suppression between
WDM and CDM abundances correlates with the warmth of the WDM models, as
indicated by the convergence of the curves to three velocity function ratios at $\vmax = 40 \, \kms$ (with the exception of the S220 model). There are some variations 
between the velocity functions of the two sterile neutrino models and their thermal best-fit model at intermediate scales
between $45 \, \kms < \vmax < 60 \, \kms$. This difference is on the
$5\%-10\%$ level, and is likely not differentiable by future high-redshift galaxy
surveys. It does suggest, however, that early stages of halo formation may have
a relevant differences that could impact the quasi-linear regime, which we explore in the 
next section.

The agreement of abundances of field halos in the thermal and sterile neutrino
WDM models, where suppression of low-mass halo formation is significant relative
to CDM, is similar to the abundance results of Local Group halos shown in Figure
\ref{fig:host_vmaxfunc}. Differences in the shape of the linear matter power
spectrum between thermal and sterile neutrino WDM models are washed out in
the deeply nonlinear regime, even at high redshift.

\section{Quasi-Linear Regime}
\label{sec:QuasiLinear}

The progression from the significant differences between thermal WDM and
resonantly-produced sterile neutrino WDM models in the linear matter power
spectrum, that sets the initial conditions for numerical simulations, to minimal
observable differences in the nonlinear measures we have explored at redshift
zero motivates the study of an intermediate, quasi-linear regime. We consider
growth of structure in the quasi-linear regime by determining the 3D nonlinear
matter power spectrum of the homogeneously-resolved cosmological
simulations at $z \sim 3-6$ described in the previous section. \cite{Viel2012} did convergence
testing for {\it N}-body simulations with the same cosmology, simulation box size, and
number of particles, and found their power spectra converged to $k = 50\,\hmpcinv$.
We therefore set $k = 50\,\hmpcinv$ as our limiting scale.

Figure \ref{fig:fullbox_ps} shows the ratio of each WDM model's matter power
spectrum to the CDM matter power spectrum at redshift $z = 2.98$ (solid) and $z = 5.429$ (dashed) in the left-hand
column. Each row in Figure \ref{fig:fullbox_ps} gives the results for a set of
WDM models of comparable warmth that includes two sterile neutrino models (one
for each V15 and
A14) and their thermal equivalent WDM model. The thin, solid
curves in each panel of the left hand column show the linear matter power
spectrum ratios for each WDM model relative to CDM. By redshift $z = 5.429$, the
nonlinear matter power spectrum calculated from the simulations have diverged
from the linear matter power spectrum by at least a few percent at
$k = 5\,\hmpcinv$ for all but the coldest WDM models (top left panel), and all
WDM models show significant evolution from the linear matter power spectrum on
scales of $k = 10\,\hmpcinv$. \citet{Viel2013} find that a few percent suppression 
in the $m = 2$~keV thermal WDM 3D nonlinear matter power spectrum relative to CDM at $k = 10\,\hmpcinv$ is differentiable in the Lyman-$\alpha$ forest 1D flux power spectrum at these redshifts. The scales and the degree of power spectrum
suppression of the sterile neutrino models relative to CDM that we find here is, therefore, relevant for Lyman-$\alpha$
forest measurements \citep{Viel2012,Viel2013}. 

The right-hand column of Figure \ref{fig:fullbox_ps} shows the power spectrum
ratios of the sterile neutrino models with respect to the matter power spectrum
of the equivalent thermal WDM model for $z = 2.98$ (solid) and $z = 5.429$ (dashed). 
The S229 and S208 model power spectrum ratios show similar departures 
from thermal WDM for $k > 2\,\hmpcinv$ at both $z = 2.98$ and $z = 5.429$. 
Compared to these two warmer V15 models, the S220 model (top right panel)
has a slightly greater reduction in power spectrum relative to thermal WDM at $k = 50\,\hmpcinv$. 
The A14 models show a greater variation in the power spectrum ratios with their thermal equivalent 
models than the qualitatively uniform V15 models. The coldest A14 model, L46, is consistent with its thermal equivalent on all scales and redshifts. 
The L7 model (middle right panel) shows suppression relative to the power spectrum of the $m = 2$~keV thermal WDM model for scales $k > 1\,\hmpcinv$. The warmest A14 model, L8 (shown in the bottom right panel), 
has more power relative to thermal WDM for $1 < k < 5(10)\,\hmpcinv$ at $z = 5.429 (2.98)$, and increasingly less power than thermal WDM on smaller scales. 

The relative strength of the resonant sterile neutrino matter power spectrum at a
fixed scale compared to a thermal equivalent WDM model does not correlate with
warmth or the details of the resonant sterile neutrino production
calculation, i.e. V15 and A14 models. For example, the L7 model shows enhanced suppression 
relative to the S229 model, but the L8 model has more power than the S208 model for $k < 50\,\hmpcinv$. 
The detailed shape of transfer function is, therefore, the most relevant 
parameter for determining the relative power in the quasi-linear regime for the scales 
probed here. 

A direct comparison with measurements of the Lyman-$\alpha$ forest is
beyond the scope of this paper. However, if a few percent difference in the 3D nonlinear matter power for 
$k = 10\,\hmpcinv$ at $z = 3-5$ can distinguish thermal WDM from CDM, as claimed \citep{Viel2012,Viel2013}, then 
the percent level differences of sterile neutrino models from thermal WDM that we find here are significant. A full Lyman-$\alpha$ study should then be able to differentiate between sterile neutrino models and a thermal equivalent model, where linear matter power spectrums differ at a few percent level on scales above the half-mode suppression scale. Constraints on WDM particle masses for models with non-thermal distributions, such as sterile neutrinos, from thermal WDM tests are therefore inadequate. Future Lyman-$\alpha$ observations that can probe scales of $10 < k < 50\,\hmpcinv$ would have great ability to test these models further.

\section{Discussion and Conclusions} 
\label{sec:Conclusions}

In this paper, we have presented a variety of simulations -- both zoom-ins of
Local Group analogs and homogeneously-resolved cosmological volumes -- of
sterile neutrino dark matter. The focus of our work is on
resonantly-produced sterile neutrinos, which result in transfer
functions that differ significantly from the standard assumption of a thermal
relic. We use transfer functions from both A14 and
V15, the latter being a more accurate calculation of the
former, 
and best-fit thermal WDM approximations. This allows a detailed
exploration of transfer function shape on Local Group halo properties while
evaluating the resonant sterile neutrino models that are best able to explain
the $3.55$~keV anomalous X-ray signal found by various surveys
\citep{Bulbul2014,Boyarsky2014}.

We find the density profiles, shapes, and mass assembly histories of the Local
Group host halo pairs are similar for all DM models, irrespective of the warmth
of the WDM model.  The abundance of low-mass subhalos in each WDM host is
reduced significantly relative to CDM and depends on the warmth of the model.
The reduction in the number of low-mass subhalos alleviates the Missing
Satellites problem. However, the number of Local Group subhalos could also be used to
rule out these and similar sterile neutrino models if the number of MW and M31
satellite galaxies exceeds the predicted number of WDM halos. This depends on
the total number of dwarf galaxies found by future surveys \citep{Tollerud2008, Hargis2014}
and the minimum DM halo mass that would host a detectable satellite galaxy
\citep{Koposov2008, Polisensky2011, Kennedy2014, Horiuchi2014}. The abundance of subhalos at the
resolution limit of the warmest sterile neutrino model considered here, L8, is
near the current satellite count for both the MW and M31 and could be ruled out in the near future 
by DES, LSST or WFIRST \citep{Ivezic2008, Bechtol2015, Spergel2015}. The abundance of low-mass
field halos is also suppressed relative to CDM, with a suppression that is
well-described by the thermal-equivalent mass of the WDM model.

The internal properties of WDM Local Group field halos and MW/M31 subhalos also 
show differences relative to their CDM counterparts: the WDM halos have a reduced central density that
becomes more pronounced for lower halo masses. However, even WDM halos several times more massive than the WDM model half-mode mass, below which free-streaming effects are typically considered most relevant, 
  show a $40\%$ decrease, on average, in their central
density halos relative to CDM. A similar result was found for subhalos in thermal WDM simulations of MW-sized DM hosts \citep{Polisensky2014,Schneider2014} and field halos in simulations using thermal approximations to resonant sterile neutrino WDM models \citep{Bose2015}. This is unsurprising, as we find the median density profile ratios and rotation curves do not show any significant variation between thermal and resonant sterile neutrino WDM models. 

Importantly, we find that the half-mode mass is largely predictive of the
number of Local Group halos, the abundance of low-mass halos
in the field, and for the density reduction in WDM halos relative to
CDM. For the range of scales probed here, nonlinear evolution washes
away linear matter power spectrum differences other than overall suppression of
small-scale power that is effectively captured by a single parameter, the
half-mode mass.  However, the more accurate resonant sterile neutrino transfer function
should be used when testing sterile neutrino models with observed
satellite galaxy counts, as differences at the 10\% level may persist and are
important in constraining dark matter models via the luminosity function of
Local Group satellites \citep{Horiuchi2015}. 
  
At high redshift, in the quasi-linear regime on scales that are relevant for
Lyman-$\alpha$ forest measurements, we find quantitative differences in the 3D
matter power spectrum of resonantly-produced sterile neutrino and thermal WDM models. The
differences in the linear matter power spectrum between the resonant sterile
neutrino DM models (based on two separate calculations by
V15 and A14) and their thermal equivalent 
WDM model are preserved in the redshift $z = 3-5$ matter power spectrum of these
models at a few percent level on small scales. The 3D matter power spectrum
differences found here could translate into relevant differences in the Lyman-$\alpha$
forest 1D flux power spectrum \citep{Viel2013}. Measurements in the
Lyman-$\alpha$ forest may be able to discern between structure formation
resulting from different sterile neutrino DM models and thermal WDM models before the epoch where nonlinear evolution washes away differences in the linear matter power spectrum.

We find that the exact transfer function shape is important for placing
constraints on resonantly-produced sterile neutrino WDM models. Thermal WDM
approximations are sufficient for some studies, such as the internal properties
of Local Group halos, and abundance of halos in the Local Group or 
large cosmological volumes. However, careful consideration should be given to constraints when comparing Local Group
luminosity functions to subhalo mass functions \citep{Horiuchi2015} and when comparing 
3D matter power spectrum results from cosmological volume simulations to
 Lyman-$\alpha$ forest measurements. Constraints on resonantly-produced sterile neutrino DM models derived from thermal WDM studies could prove out, however, simple extrapolations of those results to resonantly-produced sterile neutrino WDM models have been shown here to be incorrect. Future surveys of the
Local Group by LSST, DES, and WFIRST, as well as detailed studies of the
Lyman-$\alpha$ forest that consider scales in the quasi-linear regime,
have great power to test resonant sterile neutrino models.\\[1cm]

\section{Acknowledgments}
We thank Tejaswi Venumadhav and Francis-Yan Cyr-Racine for useful discussions. 
The authors acknowledge the University of Maryland supercomputing
resources (http://www.it.umd.edu/hpcc) made available for conducting the
research reported in this paper. MB-K acknowledges support from NASA through
Hubble Space Telescope theory grants (programs AR-12836 and AR-13888) from the
Space Telescope Science Institute (STScI), which is operated by the Association
of Universities for Research in Astronomy (AURA), Inc., under NASA contract
NAS5-26555. Support for S.G.K. was provided by NASA through Einstein Postdoctoral Fellowship grant number PF5-160136 awarded by the Chandra X-ray Center, which is operated by the Smithsonian Astrophysical Observatory for NASA under contract NAS8-03060. K.N.A. is partially supported by NSF CAREER Grant
No. PHY-11-59224 and NSF Grant No. PHY-1316792.

\bibliography{sterile_neutrino_bib_v1.bib}
\bibliographystyle{mnras.bst}
\label{lastpage}
\end{document}

%% file: extra_preamble.tex
\makeatletter
\newcommand*{\rom}[1]{\expandafter\@slowromancap\romannumeral #1@}
\makeatother

\newcommand{\vmax}{V_{\rm{max}}}

\newcommand{\mstar}{M_{\star}}
\newcommand{\msun}{M_{\odot}}
\newcommand{\hmsun}{h^{-1} \, M_{\odot}}

\newcommand{\hmpc}{h^{-1} \, {\rm Mpc}}
\newcommand{\hkpc}{h^{-1} \, {\rm kpc}}
\newcommand{\hpc}{h^{-1} \, {\rm pc}}
\newcommand{\hmpcinv}{h \, {\rm Mpc^{-1}}}

\newcommand{\kms}{{\rm km \, s}^{-1}}

\newcommand{\lcdm}{$\Lambda$CDM}

\newcommand{\mhalo}{M_{\rm halo}}